\documentclass[11pt]{article}
\pdfoutput=1
\usepackage{soul}
\usepackage{ragged2e}
\usepackage{jheppub}
\usepackage{amsfonts}
\usepackage{amsthm}
\allowdisplaybreaks[4]   
\usepackage{amsmath}
\usepackage{amssymb}   
\usepackage{placeins}
\usepackage{euscript}   
\usepackage{cleveref}    
\usepackage[dvipsnames]{xcolor}          
\usepackage{tensor}     
\usepackage{lipsum}  
\usepackage{graphicx}
\usepackage{caption}
\usepackage{afterpage}
\usepackage{subcaption}   
\usepackage{hyperref}
\usepackage{enumerate}
\hypersetup{
	colorlinks,
	linkcolor =NavyBlue,
	citecolor=MidnightBlue,
	urlcolor=NavyBlue,
}

\newcommand{\req}[1]{(\ref{#1})} 

\newcommand{\bea}{\begin{eqnarray}}
\newcommand{\eea}{\end{eqnarray}}
\newcommand{\ba}{\begin{eqnarray}}
\usepackage{braket}
\newcommand{\ea}{\end{eqnarray}}

\newcommand{\beq}{\begin{equation}}
\newcommand{\eeq}{\end{equation} }
\newcommand{\beqa}{\begin{eqnarray}}

\newcommand{\eeqa}{\end{eqnarray}}
\newcommand{\beqar}{\begin{eqnarray*}}
\newcommand{\eeqar}{\end{eqnarray*}}
\newcommand{\e}[1]{\text{e}^{#1}}

\newcommand{\be}{\begin{equation}}
\newcommand{\ee}{\end{equation}}
\newcommand{\diff}{\mathrm{d}}
\newcommand{\GN}{G_{\text N}}

\renewcommand{\req}[1]{(\ref{#1})}

\newcommand{\dd}{\mathrm{d}}

\newcommand{\iu}{\text{i}} 
\newcommand{\SLdR}{\text{SL}(2,\mathbb{R})}

\definecolor{shadecolor}{rgb}{.25,.25,.25}

\usepackage{caption}
\usepackage{subcaption}

\usepackage{multirow}
\usepackage{tikz}
\usetikzlibrary{decorations.markings,decorations.pathmorphing}
\tikzstyle{singularity}=[carmine,line width=0.5mm,decorate, decoration={zigzag,amplitude=2,segment length=6.17}]

\definecolor{carmine}{rgb}{0.59, 0.0, 0.09}
\definecolor{egyptianblue}{rgb}{0.06, 0.2, 0.65}
\definecolor{frenchlilac}{rgb}{0.53, 0.38, 0.56}
\definecolor{darkspringgreen}{rgb}{0.09, 0.45, 0.27}
\definecolor{ochre}{rgb}{0.8, 0.47, 0.13}

\usepackage{tikz-3dplot}
\tikzstyle{particle1}=[lightseagreen,line width=0.5]
\tikzstyle{particle2}=[ochre,line width=0.5]
\definecolor{lightseagreen}{rgb}{0.13, 0.7, 0.67}
\definecolor{fandango}{rgb}{0.71, 0.2, 0.54}

\title{Logarithmic correction to the entropy of near-extremal higher-curvature  black holes}

\author[a]{Pablo Bueno,}
\author[b]{Pablo A. Cano,}
\author[c]{Robie A. Hennigar,}
\author[d]{\\Javier Moreno,}
\author[e]{Ángel J. Murcia,}
\author[a]{Guido van der Velde}

\affiliation[a]{Departament de Física Quàntica i Astrofísica, Institut de Ciències del Cosmos\\ Universitat de Barcelona, Martí i Franquès 1, E-08028 Barcelona, Spain
\vspace{0.1cm}}
\affiliation[b]{Departamento de Física, Universidad de Murcia, Campus de Espinardo, 30100 Murcia, Spain
\vspace{0.1cm}}
\affiliation[c]{Centre for Particle Theory, Department of Mathematical Sciences\\ Durham University, Durham DH1 3LE, UK\vspace{0.1cm}}
\affiliation[d]{Center for Gravitational Physics and Quantum Information, \\
Yukawa Institute for Theoretical Physics, Kyoto University,\\
Kitashirakawa Oiwakecho, Sakyo-ku, Kyoto 606-8502, Japan
\vspace{0.1cm}}
\affiliation[e]{Department of Applied Mathematics and Theoretical Physics,
University of Cambridge, \\ Wilberforce Road, Cambridge CB3 0WA, United Kingdom}

\date{\today}
\abstract{
The spherically symmetric sector of broad classes of $D$-dimensional gravitational theories can be effectively described by general two-dimensional Horndeski theories. Exploiting this correspondence, we study the low-temperature dynamics of near-extremal black hole solutions of the parent theories by deriving a universal near-horizon effective description in terms of Jackiw--Teitelboim (JT) gravity. We explicitly show that the logarithmic quantum correction to the entropy, $S_{\rm JT}=\frac{3}{2}\log \left(T/T_{\rm breakdown} \right)$, 
previously derived for near-extremal black holes in Einstein gravity coupled to matter from the one-loop exact JT partition function, is universal across all models admitting an effective two-dimensional Horndeski description. We determine the general form of the scale at which this correction becomes dominant, 
$T_{\rm breakdown}$, in terms of the data of the $D$-dimensional theories.  
We illustrate our general results with charged black holes in Lovelock gravities and regular black holes in Quasitopological gravities.
}

\begin{document} 
\noindent\flushright
{YITP-26-120} 
\noindent\flushleft\justifying
\vspace{-2cm}
\vspace*{1cm}

\maketitle

\flushbottom


\section{Introduction}

Black hole thermodynamics provides one of the deepest windows into the quantum nature of gravity~\cite{Bekenstein:1972tm,Hawking:1975vcx,Bekenstein:1973ur,Hawking:1976ra,Bardeen:1973gs}. According to the \emph{central dogma}, black holes behave as ordinary quantum systems and therefore admit a statistical description in terms of an underlying microscopic Hilbert space~\cite{Almheiri:2020cfm}. 
Understanding how this description emerges remains one of the central challenges in quantum gravity. 
In this context, the Euclidean gravitational path integral~\cite{Gibbons:1976ue} has proven to be a surprisingly powerful tool, reproducing the classical laws of black hole thermodynamics~\cite{Gibbons:1976pt,Hawking:1982dh,York:1986it,Hawking:1995fd}, systematically capturing universal quantum corrections~\cite{Banerjee:2010qc,Banerjee:2011jp,Sen:2012cj,Ghosh:2019rcj,Heydeman:2020hhw,Iliesiu:2020qvm,Iliesiu:2022onk,Boruch:2022tno,Kolanowski:2024zrq}, 
accounting for the Page curve in black hole evaporation~\cite{Penington:2019npb,Penington:2019kki,Almheiri:2019qdq}, and providing a statistical-mechanical understanding of black hole entropy in increasingly broad settings~\cite{Balasubramanian:2022gmo,Climent:2024trz}.

Near-extremal black holes provide a particularly interesting arena in which to explore these questions. On the one hand, they probe the regime in which quantum effects become increasingly important. On the other hand, it has long been appreciated that the conventional thermodynamic description cannot remain valid arbitrarily close to extremality~\cite{Preskill:1991tb,Maldacena:1998uz}. Indeed, a thermal treatment is reliable only when the emission
of a typical Hawking quantum, carrying energy of order $T$, produces a negligible change in the black-hole temperature. This condition inevitably fails at sufficiently low temperatures.

A universal feature of stationary extremal black holes is the emergence of an infinitely long near-horizon throat. For static, spherically symmetric  (SS) solutions, this geometry contains an AdS$_2$ spacetime fibered over the transverse S$^{D-2}$ directions. The large separation between the throat and the asymptotic region provides a natural decomposition of the full quantum system into infrared and ultraviolet degrees of freedom: the near-horizon throat captures the relevant low-energy degrees of freedom, while the far region acts as the external environment~\cite{Nayak:2018qej}.  Such a splitting makes it possible to investigate universal aspects of the quantum dynamics of near-extremal black holes without requiring a complete microscopic theory of quantum gravity. This has been explicitly established for charged black holes in Einstein gravity in three~\cite{Ghosh:2019rcj,Maxfield:2020ale}, four~\cite{Iliesiu:2020qvm,Iliesiu:2022onk}, and higher dimensions~\cite{Boruch:2022tno}; for near-Nariai black holes, where the relevant low-energy limit is associated with the coincidence of the black-hole and cosmological horizons~\cite{Maldacena:2019cbz,Maulik:2025phe,Blacker:2025zca}; for rotating black holes in theories with an Einstein-Hilbert gravitational sector and general matter couplings~\cite{Kapec:2023ruw,Rakic:2023vhv,Maulik:2024dwq,Kapec:2024zdj,Modak:2025gvp,PandoZayas:2026vbg}; and for several modified-gravity theories in various dimensions~\cite{Banerjee:2021vjy,Rathi:2021aaw,Alvarado:2026kio,Acito:2026mmf,Despontin:2026xzg}.

In these cases, it was shown that the infrared dynamics is universally controlled by the emergent $\SLdR$ symmetry of the AdS$_2$ near-horizon region~\cite{Maldacena:2016upp}, which becomes exact in the extremal limit. Near extremality, the explicit breaking of this symmetry gives rise to boundary reparametrization modes that become parametrically light and therefore remain dynamical in the low-energy effective theory~\cite{Maldacena:2016upp,Stanford:2017thb,Saad:2019lba}. These modes produce universal quantum corrections to the gravitational path integral. In the SS sector, these degrees of freedom correspond to fluctuations of the AdS$_2$ metric together with a dilaton-like mode, which measures the area of the transverse sphere. Their dynamics is universally described by Jackiw--Teitelboim (JT) gravity~\cite{Jackiw:1984je,Teitelboim:1983ux}, which therefore provides the effective theory governing the low-temperature dynamics of near-extremal black holes~\cite{Gaikwad:2018dfc,Nayak:2018qej,Moitra:2019bub,Sachdev:2019bjn,Mertens:2022irh}. A remarkable feature of JT gravity is that its perturbative genus-zero path integral localizes onto the Schwarzian boundary mode~\cite{Ghosh:2019rcj} and is consequently one loop exact~\cite{Stanford:2017thb}. As a result, the temperature-dependent quantum correction to the near-extremal black-hole entropy is universal,~\cite{Turiaci:2023wrh}
\begin{equation}\label{JTuni}
S_{\rm JT}=\frac{3}{2}\log\frac{T}{T_{\rm breakdown}}\,,
\end{equation}
where $T_{\rm breakdown}$ is a theory-dependent scale that sets the normalization of the logarithmic correction and parametrically marks the onset of the regime in which it competes with the semiclassical linear-in-$T$ contribution.

In this paper, we show that this universal logarithmic correction to the entropy, \req{JTuni}, arises for general static and spherically symmetric (SS) near-extremal black holes in generic $D$-dimensional higher-curvature theories of gravity whose SS sector is described by two-dimensional Horndeski gravities.\footnote{Quantum matter can also induce logarithmic modifications of black-hole geometries through semiclassical backreaction, particularly in higher-curvature theories~\cite{Martinez:1996uv,Casals:2016odj,Emparan:2020znc,Battista:2023iyu,Chernicoff:2024dll,Wang:2025fmz,Frassino:2024bjg,Climent:2024wol,Mendez-Zavaleta:2026rgg}.} The correction we study here is of a purely gravitational origin: it follows from the one-loop exactness of the Schwarzian/JT sector governing the near-horizon throat, and is present already in the absence of additional matter fields. In particular, our results apply to black hole solutions of Lovelock~\cite{lovelock1970divergence,Lovelock:1971yv} and Quasitopological gravities~\cite{Oliva:2010zd,Myers:2010ru,Bueno:2025qjk,Dehghani:2011vu, Cisterna:2017umf, Ahmed:2017jod,Bueno:2019ycr, Bueno:2022res, Moreno:2023rfl,Moreno:2023arp}, including infinite families of pure-gravity theories whose only SS solutions are regular black holes~\cite{Bueno:2024dgm,Bueno:2025zaj,Aguayo:2025xfi,Hennigar:2025yqm,Borissova:2026wmn,Bueno:2026oyg}.\footnote{Regular black holes also arise in Electromagnetic Quasitopological gravities, in which higher-curvature terms are non-minimally coupled to electromagnetic degrees of freedom~\cite{Cano:2020ezi,Cano:2020qhy,Bueno:2021krl,Bueno:2022ewf,Bueno:2025dqk}, but we shall not consider those cases here.} More generally, every two-dimensional Horndeski theory can be obtained by the spherical reduction of a suitable---possibly non-polynomial---$D$-dimensional higher-curvature gravity theory~\cite{Borissova:2026krh}. Our results therefore show that the universality of this logarithmic correction to the entropy, Eq.~\eqref{JTuni}, extends to an extremely broad class of higher-curvature theories of gravity.

The paper is organized as follows. In Section~~\ref{2dH} we review two-dimensional Horndeski theories of gravity and write down the general equations satisfied by the two independent physical degrees of freedom they describe. We also derive general formulas for the temperature, Wald entropy, and mass of the would-be near-extremal black holes associated with these theories. In Section~\ref{hdo2} we explain how the SS sectors of various $D$-dimensional higher-curvature theories of gravity reduce to two-dimensional Horndeski theories. We explicitly perform the dimensional reduction for Einstein and Lovelock gravities, as well as for Quasitopological theories, and then, following~\cite{Borissova:2026krh}, we show how any two-dimensional Horndeski theory arises as the SS reduction of at least one $D$-dimensional theory of gravity. In Section~\ref{sec:low temperature dynamics} we exploit the effective Horndeski description to study the low-temperature dynamics of general SS $D$-dimensional black holes described by these theories. We show that the semiclassical thermodynamic properties receive a universal quantum correction controlled by an effective JT partition function which we evaluate explicitly. The result is a universal modification to the black hole entropy of the form~\eqref{JTuni} which dominates at low temperatures. The breakdown temperature is determined as a function of the couplings of the higher-dimensional theories. In Section~\ref{examp} we exemplify our result in the case of charged black holes in Lovelock gravity as well as for pure gravity regular black holes in Quasitopological theories. We conclude in Section~\ref{conclu}.

\section{Two-dimensional Horndeski gravities}\label{2dH}

Let us consider the most general two-dimensional theory for a scalar field $\varphi$ and a two-dimensional Lorentzian metric $\gamma=\gamma_{\mu \nu} \mathrm{d}x^\mu \mathrm{d}x^\nu$ with second-order equations of motion,
\begin{equation}
\mathcal{L}_{\rm 2d}=G_2(\varphi,X)-G_3(\varphi,X)\Box \varphi +G_4(\varphi,X)R-2 G_{4,X}(\varphi,X) \left[(\Box \varphi )^2-\nabla_\mu\nabla_\nu\varphi\nabla^\mu\nabla^\nu\varphi\right]\, ,
\label{eq:horngen}
\end{equation}
where $G_i(\varphi,X)$ for  $i={2,3,4}$ correspond to arbitrary functions of $\varphi$ and $X\equiv(\partial \varphi)^2$, $G_{4,X}$ is the partial derivative with respect to $X$ of $G_4$ and $R$ is the Ricci scalar of $\gamma$. These theories are the two-dimensional analogues of the most general scalar-tensor four-dimensional theories with second-order equations of motion found by Horndeski~\cite{Horndeski:1974wa,Kobayashi:2011nu,Kobayashi:2019hrl}. Their equations of motion read~\cite{Borissova:2026krh, Carballo-Rubio:2025ntd}
\begin{eqnarray}\label{eomg}
\mathcal{E}_{a b}&=&\beta\nabla_a\nabla_b \varphi-g_{ab}\left(\frac{1}{2}\alpha+\beta\Box \varphi\right)+\left(\partial_X\alpha-\partial_\varphi \beta\right)\nabla_a\varphi\nabla_b \varphi\, ,\\
\mathcal{E}_\varphi&=& -\beta R+2\partial_\varphi\beta\Box \varphi+\partial_\varphi\alpha+2\partial_X\beta\left[(\Box\varphi)^2-\nabla_a\nabla_b\varphi\nabla^a\nabla^b\varphi\right]\\
&&-2\partial_\varphi\left(\partial_X\alpha-\partial_\varphi\beta\right)X-2\left(\partial_X\alpha-\partial_\varphi\beta\right)\Box \varphi-2\partial_X\left(\partial_X\alpha-\partial_\varphi\beta\right)\nabla_a \varphi\nabla^a X\, ,
\end{eqnarray}
where we defined
\begin{equation}\label{alpha_beta}
\alpha\equiv G_2+X\partial_\varphi\left(G_3-2\partial_\varphi G_4\right)\, , \quad \beta\equiv X\partial_X\left(G_3-2\partial_\varphi G_4\right)-\partial_\varphi G_4\, .
\end{equation}
In principle, the solutions to these theories are described by four independent functions, namely, the components of the two-dimensional metric $\gamma_{tt}=\gamma_{tt}(t,r)$, $\gamma_{tr}=\gamma_{tr}(t,r)$, $\gamma_{rr}=\gamma_{rr}(t,r)$ and the scalar field $\varphi=\varphi(t,r)$. However, invariance under general coordinate transformations is parametrized by two arbitrary functions and thus gauge-fixing removes two of them.\footnote{ Equivalently, varying the action along a diffeomorphism generated by an arbitrary vector and using invariance of the action yields the Bianchi identity
\begin{equation}
\nabla^a\mathcal{E}_{a b}+\frac{1}{2}\mathcal{E}_\varphi\nabla_b\varphi=0\, .
\end{equation}
} In particular, we fix $\varphi=r$ and $\gamma_{tr}=0$. After relabeling $\gamma_{tt}\equiv -N(t,r)^2f(t,r)$ and $\gamma_{rr}\equiv 1/f(t,r)$, the most general metric solving the equations of motion of theory~\eqref{eq:horngen} can be written as
\begin{equation}\label{eq:metrictr}
\diff s_{\rm 2d}^2=-N(t,r)^2 f(t,r)\diff t^2+ \frac{\diff r^2}{f(t,r)}\,.
\end{equation}

From fixing $\varphi=r$ we  see that $X=f$. From the off-diagonal component of the equations of motion~\eqref{eomg} we observe that $\partial_t f=0$ and hence $f=f(r)$. This allows us to remove the time dependence of $N$ by a redefinition of $t$. The remaining two components, $\mathcal{E}_{tt}=0$ and $\mathcal{E}_{rr}=0$, then give
\begin{equation}\label{eomf}
\alpha+\beta f'=0\,,
\end{equation}
\begin{equation}\label{eomN}
\frac{\partial_r N}{N}=\frac{\partial_f \alpha-\partial_r\beta}{\beta}\,,
\end{equation}
where we denoted $f'=\diff f/\diff r$. Eq.~\eqref{eomN} can be integrated to obtain
\begin{equation}\label{N(r)}
N(r)=\exp{\left\{\int_{\infty}^{r}\diff r' \frac{\partial_f \alpha-\partial_{r'}\beta}{\beta}\right\}}\, . 
\end{equation} 
Consequently, up to the residual freedom to redefine the time coordinate, the solutions in the gauge $\varphi=r$ are static. In the following, we therefore consider the line element
\begin{equation}\label{eqr}\diff s_{\rm 2d}^2=-N(r)^2 f(r)\diff t^2+\frac{\diff r^2}{f(r)}\,.\end{equation}

In the following sections we will consider higher-dimensional gravitational theories which admit (near)-extremal black-hole solutions and whose spherically symmetric (SS) sector is captured by two-dimensional Horndeski gravities. 
Using Eq.~(\ref{eomf}), we may write the temperature of such black holes  as 
\begin{equation}
T=\frac{1}{4\pi}N(r_+)f'(r_+)=\frac{1}{4\pi} \exp{\left\{\int_{\infty}^{r_+}\diff r \frac{\partial_f \alpha-\partial_r\beta}{\beta}\right\}}\frac{G_2(r_+,0)}{ \partial_{r_+}G_4(r_+,0)}\, ,
\end{equation}
where $\alpha$ and $\beta$ are defined in Eq.~\eqref{alpha_beta}, and $r_+$ is the horizon radius $f(r_+)=0$. Furthermore, the Wald entropy~\cite{Wald:1993nt,Iyer:1994ys} reads
\begin{equation}\label{Wald_entropy}
S=\frac{(D-2)\Omega_{D-2}}{4 \GN}G_4(r_+,0)\, .
\end{equation}
Hence, the first law of thermodynamics gives for the mass
\begin{equation}
M=\int \diff r_+ T \partial_{r_+}S= \frac{(D-2)\Omega_{D-2}}{16\pi \GN}\int \diff r_+ \exp{\left\{\int_{\infty}^{r_+}\diff r \frac{\partial_f \alpha-\partial_r\beta}{\beta}\right\}} G_2(r_+,0)\, .
\end{equation}
Every higher-dimensional gravity admitting a spherical reduction to a two-dimensional Horndeski theory will be characterized by certain functions $G_{i}$ determined by  the corresponding higher-dimensional couplings. The above formulas capture the thermodynamic properties of the SS black hole solutions of every such theory---see also \cite{Borissova:2026rbi}.

\section{Higher-dimensional origins of two-dimensional Horndeski gravities}\label{hdo2}

Recently, it has been shown that a large subclass of two-dimensional Horndeski theories of the form~\eqref{eq:horngen} arise as the spherical reduction of special classes of $D\geq 5$ higher-curvature gravities known as \emph{Quasitopological gravities}~\cite{Bueno:2024eig,Bueno:2024zsx}, as well as of $D=4$ non-polynomial theories of the same type~\cite{Bueno:2025zaj}. This was found to be particularly relevant, as these theories naturally admit regular black holes as their unique SS solutions~\cite{Bueno:2024dgm}. Remarkably, such regular black holes are formed dynamically from the collapse of minimally coupled matter~\cite{Bueno:2024eig,Bueno:2024zsx,Bueno:2025gjg}.

These results triggered the following question: given any two-dimensional Horndeski theory of the form~\eqref{eq:horngen}, is it always possible to find one, or several, higher-dimensional $(D\geq 4)$ theories of gravity whose SS sectors are captured by it?
Interestingly enough, the answer turns out to be 
positive~\cite{Colleaux:2017ibe,Colleaux:2019ckh,Borissova:2026wmn,Borissova:2026krh}. We devote this section to reviewing this construction. In particular, we will see that the mapping is in general surjective but non-injective, namely, one can always find at least one higher-curvature theory whose SS sector is described by a given two-dimensional Horndeski theory, but such higher-dimensional theory is usually non-unique.

Let $(M,g)$ be a $D$-dimensional Lorentzian manifold with $D \geq 4$ and let  $\mathsf{R}_{ab}{}^{cd}$, $\mathsf{W}_{ab}{}^{cd}$, $\mathsf{Z}_{a}{}^b$ and $\mathsf{R}$ be the Riemann curvature tensor, the Weyl curvature tensor, the traceless Ricci tensor and the Ricci scalar of $g$, respectively. We will assume that $(M,g)$ decomposes as the following warped product: 
\begin{equation}
   (M,g)=\left (M_2 \times \rm S^{D-2},\text{ } \mathrm{d}s^2_{\gamma}=\gamma_{\mu \nu} \mathrm{d}x^\mu \mathrm{d}x^\nu +\varphi^2(x) \mathrm{d}\Omega_{D-2}^2\right)\,, 
       \label{eq:sphermet}
\end{equation}
where $(M_2,\gamma_{\mu \nu} \mathrm{d}x^\mu \mathrm{d}x^\nu)$ is a two-dimensional Lorentzian manifold and $\mathrm{d}\Omega_{D-2}^2$ denotes the metric of the round sphere $\rm S^{D-2}$. Observe that Greek indices $\{\mu, \nu, \dots\}$ are used for tensors living on $M_2$ and Latin indices $\{a,b,\dots\}$ for tensors on $M$.  Evaluated on Eq.~\eqref{eq:sphermet}, the Weyl tensor, traceless Ricci tensor and Ricci scalar are given by:
\begin{align}
\label{eq:weyldec}
\left. \mathsf{W}_{ab}{}^{cd} \right \vert_\gamma&=\Omega \left[\frac{(D-2)(D-3)}{2} \gamma_{[a}{}^{c} \gamma_{b]}{}^{d}+\sigma_{[a}{}^{c} \sigma_{b]}{}^{d} -(D-3) \gamma_{[a}{}^{[c} \sigma_{b]}{}^{d]} \right] \,, \\
\label{eq:riccidec}
\left. \mathsf{Z}_{ab} \right \vert_\gamma&=\delta_{a}^\mu \delta_{b}^\nu \left (  \Xi \gamma_{\mu \nu}  -(D-2)\frac{\nabla_\mu \nabla_\nu \varphi}{\varphi}\right)+\Theta \, \sigma_{ab}\,,  \\
\label{eq:scalD}
\left. \mathsf{R}^{(D)} \right \vert_\gamma&=R-(D-2) \left[ \frac{2\Box \varphi}{\varphi}-(D-3) \psi \right] \,,
\end{align}
where $\gamma_{ab}=\delta_{a}^\mu \delta_{b}^\nu \gamma_{\mu \nu}$, $\sigma_{ab}=g_{ab}-\gamma_{ab}$, $R$ stands for the Ricci scalar of $\gamma_{\mu \nu}$, $\Box \varphi$ is the Laplacian of $\varphi$ associated with $\gamma_{\mu \nu}$ and:
\begin{align}
\label{eq:psiXdef}
\psi&\equiv \frac{1-X}{\varphi^2}\, ,\quad X\equiv\nabla_{\mu}\varphi\nabla^{\mu}\varphi\, ,  \quad 
\Omega\equiv\frac{2(2\varphi^2 \psi+2 \varphi \Box \varphi +\varphi^2 R)}{(D-1)(D-2) \varphi^2}\,,\\
\Theta&\equiv\frac{2(D-3)\varphi^2 \psi+(D-4) \varphi \Box \varphi-\varphi^2 R}{D \varphi^2}\,, \quad
\Xi\equiv \frac{(D-2)}{D}\left (\frac{R}{2}+\frac{2 \Box \varphi}{\varphi}-(D-3) \psi \right)\,.
\end{align}
Given a certain $D$-dimensional theory of gravity\footnote{We use the same notation for the Levi-Civita covariant derivative associated with $(M,g)$ and for that of $(M_2,\gamma_{\mu \nu} \mathrm{d}x^\mu \mathrm{d}x^\nu)$. It will be clear from the context---and from the indices employed---which covariant derivative we are referring to in each case.}
\begin{equation}\label{eq:genDaction}
S_D= \frac{1}{16\pi G_{\rm N} }\int \mathrm{d}^Dx \sqrt{|g|}\mathcal{L}_{\rm D}\left (g^{ab}, \mathsf{R}_{abcd},\nabla_a\right )\, ,
\end{equation}
we will focus on deriving the two-dimensional theory $S_{\rm 2d}$ for a two-dimensional Lorentzian metric $\gamma_{\mu \nu}$ and a scalar $\varphi$  that arises upon the spherical reduction of~\eqref{eq:genDaction}, which will take the form
\begin{equation}\label{eq:2dgenaction}
S_{\rm 2d}=\frac{(D-2)\Omega_{D-2}}{16\pi G_{\rm N}}\int \mathrm{d}^{2}x\sqrt{|\gamma|} \mathcal{L}_{\rm 2d}(\gamma_{\mu\nu},\varphi)\, ,
\end{equation}
with  $\Omega_{(D-2)}\equiv 2\pi^{(D-1)/2}/\Gamma\left[\tfrac{D-1}{2}\right]$. In the following, we will be interested in obtaining the two-dimensional Lagrangians $\mathcal{L}_{\rm 2d}(\gamma_{\mu\nu},\varphi)$ arising from the spherical reduction of some relevant theories of gravity. 

\subsection{Spherical reduction of GR and Lovelock gravities}\label{sec:GR&Lovelock}

Let us begin with the simplest and most prominent diffeomorphism-invariant theory  of gravity, which corresponds to general relativity (GR):
\begin{equation}
    \mathcal{L}_{\rm D}\left (g^{ab}, \mathsf{R}_{abcd}\right )=\mathsf{R}\,.
\end{equation}
In such a case, direct use of~\eqref{eq:scalD} provides the following two-dimensional Horndeski theory:
\begin{equation}
\mathcal{L}_{\rm 2d}^{\rm GR}(\gamma_{\mu \nu},\varphi)=G_{2}^{\rm GR}(\varphi, X)-\Box\varphi G_{3}^{\rm GR}(\varphi, X)+G_{4}^{\rm GR}(\varphi, X)R\, ,
\end{equation}
where the various functions $G_{i}^{\rm GR}(\varphi, X)$ with $i=2,3,4$ read
\begin{equation}\label{G_GR}
G_{2}^{\rm GR}(\varphi, X)= (D-3) \varphi^{D-2} \psi\,, \quad  G_{3}^{\rm GR}(\varphi, X)= 2 \varphi^{D-3} \,, \quad G_{4}^{\rm GR}(\varphi, X)= \frac{1}{D-2}\varphi^{D-2}\,.
\end{equation}
A more intricate family of theories is given by Lovelock gravities~\cite{lovelock1970divergence,Lovelock:1971yv}. These include GR and constitute the most general theories whose equations of motion are of second-derivative order. A generic theory of this class may be written in the form
\begin{equation}
     \mathcal{L}_{\rm D}^{\rm L}\left (g^{ab}, \mathsf{R}_{abcd}\right )=\mathsf{R}+\sum_{n=2}^{\lfloor D/2\rfloor} \frac{(2n)! (D-2n)! }{2^n (D-2)! }  \alpha_ n \ell^{2n-2}  \mathsf{R}_{[a_1 a_2}^{a_1 a_2} \dots \mathsf{R}_{a_{2n-1} a_{2n}]}^{a_{2n-1} a_{2n}}\,,
     \label{eq:lovelock}
\end{equation}
where $\alpha_n$ are arbitrary dimensionless couplings, $\ell$ is a certain length scale and $\lfloor D/2\rfloor$ stands for the integer part of $D/2$. For $D=2n$, the terms of $n$-th order in curvature combine into a topological theory that does not contribute to the equations of motion. Reducing on Eq.~\eqref{eq:sphermet}, the theory~\eqref{eq:lovelock} becomes a two-dimensional Horndeski theory given by:
\begin{equation}
\label{eq:hornL}
    \mathcal{L}_{\rm 2d}^{\rm L}(\gamma_{\mu \nu},\varphi)=G_{2}^{\rm L}(\varphi, X)-\Box\varphi G_{3}^{\rm L}(\varphi, X)+G_{4}^{\rm L}(\varphi, X)R-2G_{4,X}^{\rm L}(\varphi, X)\left[(\Box\varphi)^2-\nabla_{\mu}\nabla_{\nu}\varphi\nabla^{\mu}\nabla^{\nu}\varphi\right]\, ,
\end{equation}
where now the functions $G_{i}^{\rm L}(\varphi, X)$ with $i=2,3,4$ correspond to
\begin{align}
\label{eq:G23L}
G_{2}^{\rm L}(\varphi, X)&=\varphi^{D-2}\left[(D-1)h(\psi)-2\psi  h'(\psi)\right]\, \quad 
G_{3}^{\rm L}(\varphi, X)=2\varphi^{D-3}h'(\psi)\, ,\\
\label{eq:G4L}
G_{4}^{\rm L}(\varphi, X)&=-\frac{1}{2}\varphi^{D-2}\psi^{(D-2)/2}\int \mathrm{d}\psi \,\psi^{-D/2}h'(\psi)\, ,
\end{align}
where we have implicitly defined the \emph{characteristic polynomial}
$h(\psi)$, 
 \begin{equation}
 \label{eom_psiL}
h(\psi)\equiv \psi + \sum_{n=2}^{\lfloor D/2\rfloor}\alpha_n \ell^{2n-2} \frac{D-2n}{D-2}  \psi^n\, ,
\end{equation}
and where one must choose the primitive in Eq.~\eqref{eq:G4L} such that $G_{4}^{\rm L}(1,1)=1/(D-2)$. Observe that all the information about the higher-dimensional Lovelock theory is entirely encoded in the characteristic polynomial ~\cite{Camanho:2010ru,Camanho:2011rj,Camanho:2013pzg}.

\subsection{Spherical reduction of Quasitopological gravities}

Some years ago, a new class of higher-curvature theories constructed from linear combinations of polynomial curvature invariants was identified in $D \geq 5$ spacetime dimensions~\cite{Oliva:2010zd,Myers:2010ru}. These theories were called \emph{Quasitopological gravities} (QT gravities) and are nowadays defined by the property of having second-order equations on spherically symmetric backgrounds~\cite{Bueno:2025qjk}. As such, QT gravities are understood as generalizations of Lovelock gravities and have been found at any curvature order~\cite{Dehghani:2011vu, Cisterna:2017umf, Ahmed:2017jod,Bueno:2019ycr, Bueno:2022res, Moreno:2023rfl,Moreno:2023arp}. Among their most remarkable features, QT gravities were proven to capture the vacuum effective theory of gravity~\cite{Bueno:2019ltp} and naturally admit regular black holes (when infinite towers of QT densities are included in the action)~\cite{Bueno:2024dgm} as their unique SS solutions~\cite{Bueno:2024zsx,Bueno:2024eig,Bueno:2025qjk}. 

A generic QT gravity is written as a (possibly infinite) linear combination of QT gravities of a fixed curvature order. Specifically,
\begin{equation}
     \mathcal{L}_{\rm D}^{\rm QT}\left (g^{ab}, \mathsf{R}_{abcd}\right )=\mathsf{R}+\sum_{n=2}^{\infty}  \alpha_n \ell^{2n-2}  \mathcal{Z}_{(n)}\,,
     \label{eq:genQT}
\end{equation}
where $\alpha_n$ are dimensionless couplings, $\ell$ is a length scale and $\mathcal{Z}_{(n)}$ are QT gravities constructed purely from curvature invariants of $n$-th order. Observe that the series can be truncated at any desired order $n_{\rm max}$ by setting $\alpha_{n> n_{\rm max}}=0$. Beyond quadratic order, the class of QT gravities is largely degenerate and there exist different choices of $\mathcal{Z}_{(n)}$ at each order giving rise to the same equations on spherical symmetry. We will not elaborate further on this issue and will just focus on presenting certain particular representatives at every curvature order. To this end, we find it convenient to introduce the following notation:
\begin{align}
\mathsf{W}_2&\equiv \mathsf{W}_{ab}{}^{cd} \mathsf{W}_{cd}{}^{ab}\,, \quad \mathsf{Z}_2\equiv\mathsf{Z}_{a}{}^{b} \mathsf{Z}_{b}{}^{a}\,,  \quad \mathsf{W}_3\equiv\mathsf{W}_{ab}{}^{cd} \mathsf{W}_{cd}{}^{ef} \mathsf{W}_{ef}{}^{ab}\,,
\\  \mathsf{Z}_3&\equiv\mathsf{Z}_{a}{}^{b} \mathsf{Z}_{b}{}^{c} \mathsf{Z}_{c}{}^{a}\,, \quad 
\mathsf{X}_3\equiv\mathsf{W}_{abc}{}^{d} \mathsf{W}^{abc}{}_{e} \mathsf{Z}_{d}{}^{e} \,, \quad \mathsf{Y}_3\equiv\mathsf{W}_{ab}{}^{cd} \mathsf{Z}_{c}{}^{a} \mathsf{Z}_{d}{}^{b}\,,\\
\mathsf{X}_4&\equiv\mathsf{W}_{acbd} \mathsf{W}^{c efg} \mathsf{W}^d{}_{efg} \mathsf{Z}^{ab}\,, \quad \mathsf{U}_4\equiv\mathsf{W}_{abcd} \mathsf{W}^{aecf} \mathsf{Z}^{bd} \mathsf{Z}_{ef}\,, \\ \mathsf{Y}_4&\equiv\mathsf{Z}_{a c} \mathsf{Z}_{d e} \mathsf{W}^{bdce} \mathsf{Z}^{a}_b\,, \quad \mathsf{Z}_4\equiv\mathsf{Z}_a^b \mathsf{Z}_b^c \mathsf{Z}_c^d \mathsf{Z}_d^a\,, \\
\mathsf{X}_5&\equiv\mathsf{Z}^a_b  \mathsf{Z}^b_c \mathsf{W}_{daef} \mathsf{W}^{efgh} \mathsf{W}_{gh}{}^{dc}\,, \quad \mathsf{U}_5\equiv\mathsf{Z}^{ab} \mathsf{W}_{acbd} \mathsf{Z}^{ef}  \mathsf{W}\indices{_e^c_f^g} \mathsf{Z}^d_g\,,\\
\mathsf{Y}_5&\equiv\mathsf{Z}_{a}^b \mathsf{Z}_{b}^{c} \mathsf{Z}_{cd} \mathsf{Z}_{ef} \mathsf{W}^{eafd}\,, \quad \mathsf{Z}_5\equiv\mathsf{Z}_a^b \mathsf{Z}_b^c \mathsf{Z}_c^d \mathsf{Z}_d^e \mathsf{Z}_e^a\,.
\end{align}
Defining $\mathcal{Z}_{(1)}\equiv\mathsf{R}$ for completeness, we choose the following QT gravities up to quintic order:
\begin{align}
 \mathcal{Z}_{(2)}&=\frac{1}{(D-2)} \left [\frac{\mathsf{W}_2}{D-3} -\frac{4 \mathsf{Z}_2}{D-2}\right]+\frac{\mathcal{Z}_{(1)}^2}{D(D-1)}\,, \\
\nonumber \mathcal{Z}_{(3)}&=\frac{24}{(D-2)(D-3)} \left[\frac{ \mathsf{Y}_3}{(D-2)^2}-
   \frac{ \mathsf{X}_3}{(D-2) (D-4)}+\frac{2(D-3)
   \mathsf{Z}_3}{3(D-2)^3} \right] \\&+ \frac{2(2 D-3) \mathsf{W}_3}{ (D-2)(D-3)(D^3-9 D^2+26D-22)} + \frac{3\mathcal{Z}_{(1)}\mathcal{Z}_{(2)}}{D(D-1)}-\frac{2 \mathcal{Z}_{(1)}^3}{D^2(D-1)^2} \,, \\
\nonumber
\mathcal{Z}_{(4)}&=\frac{96}{(D-2)^2(D-3)} \left[\frac{(D-1)\left ( \mathsf{W}_2 \right)^2}{8D(D-2)^2(D-3)}-\frac{(2D-3) \mathsf{W}_2\mathsf{Z}_2}{4(D-1)(D-2)^2}-
\frac{2 \mathsf{X}_4}{D(D-3)(D-4)}\right. \\ \nonumber & -\frac{4\mathsf{Y}_4}{(D-2)^2(D-4)} \left. +\frac{(D^2-3D+3) \left (\mathsf{Z}_2\right )^2}{D(D-1)(D-2)^3}-\frac{\mathsf{Z}_4}{(D-2)^3}+\frac{(2D-1)\mathsf{U}_4}{D(D-2)(D-3)}\right]\\&+\frac{4\mathcal{Z}_{(1)}\mathcal{Z}_{(3)}-3 \mathcal{Z}_{(2)}^2}{D(D-1)}\,,\\ \nonumber
\mathcal{Z}_{(5)}&=\frac{960 (D-1)}{(D-2)^4(D-3)^2} \left[ \frac{(D-2)\mathsf{W}_2\mathsf{W}_3 }{40D(D^3-9 D^2+26D-22)}+\frac{4(D-3) \mathsf{Z}_5}{5(D-1)(D-2)^2(D-4)}\right. \\ \nonumber & -\frac{(3D
-1)\mathsf{W}_2  \mathsf{X}_3}{10D(D-1)^2(D-4)}-\frac{4(D-3)(D^2-2D+2)\mathsf{Z}_2 \mathsf{Z}_3}{5D(D-1)^2(D-2)^2(D-4)} +\frac{(5D^2-7D+6)\mathsf{Z}_2 \mathsf{Y}_3}{10D(D-1)^2(D-2)}\\ \nonumber & -\frac{(D-3)(3D-1)(D^2+2D-4)\mathsf{W}_2 \mathsf{Z}_3}{10D(D-1)^2(D+1)(D-2)^2(D-4)}- \frac{2(3D-1) \mathsf{U}_5}{D(D^2-1)(D-4)}-\frac{\mathsf{Y}_5}{(D-1)(D-2)} \\ \nonumber & +\frac{(D-2)(D-3)(15D^5-148 D^4+527 D^3-800 D^2+472D-88)\mathsf{W}_3 \mathsf{Z}_2}{40D(D-1)^2(D-4)(D^5-15D^4+91 D^3-277 D^2+418D-242)}\\ \nonumber &\left.+\frac{(D-3)\mathsf{Z}_2 \mathsf{X}_3}{5D(D-1)^2(D-4)} -\frac{(D-2)(D-3)(3D-2) \mathsf{X}_5}{4(D-1)^2(D-4)(D^2-6D+11)}+\frac{ \mathsf{W}_2\mathsf{Y}_3}{20D(D-1)^2}\right]\\&+\frac{5\mathcal{Z}_{(1)}\mathcal{Z}_{(4)}-2\mathcal{Z}_{(2)}\mathcal{Z}_{(3)}}{D(D-1)}+\frac{6 \mathcal{Z}_{(1)}\mathcal{Z}_{(2)}^2-8 \mathcal{Z}_{(1)}^2\mathcal{Z}_{(3)}}{D^2(D-1)^2}\,.
\end{align}
Observe that the quadratic theory $\mathcal{Z}_{(2)}$ is just the Gauss-Bonnet Lagrangian, while the cubic theory is the lowest-order QT gravity beyond the Lovelock class. To obtain even higher-curvature orders $n$, we may use the recursion formula~\cite{Bueno:2019ycr,Bueno:2024zsx}
\begin{equation}
\label{eq:zrec}
\mathcal{Z}_{(n+5)}=\frac{3(n+3)\mathcal{Z}_{(1)}\mathcal{Z}_{(n+4)}}{D(D-1)(n+1)}-\frac{3(n+4)\mathcal{Z}_{(2)}\mathcal{Z}_{(n+3)}}{D(D-1)n}+\frac{(n+3)(n+4)\mathcal{Z}_{(3)}\mathcal{Z}_{(n+2)}}{D(D-1)n(n+1)}\, .
\end{equation}
After these specifications, we are ready to perform the spherical reduction of the theory ~\eqref{eq:genQT}. As it turns out, the following two-dimensional Horndeski theory is produced~\cite{Bueno:2024eig,Bueno:2024zsx}:
\begin{align}
\notag
    \mathcal{L}_{\rm 2d}^{\rm QT}(\gamma_{\mu \nu},\varphi)&=G_{2}^{\rm QT}(\varphi, X)-\Box\varphi G_{3}^{\rm QT}(\varphi, X)+G_{4}^{\rm QT}(\varphi, X)R\\\label{eq:hornQT}&-2G_{4,X}^{\rm QT}(\varphi, X)\left[(\Box\varphi)^2-\nabla_{\mu}\nabla_{\nu}\varphi\nabla^{\mu}\nabla^{\nu}\varphi\right]\, ,
\end{align}
where  the functions $G_{i}^{\rm QT}(\varphi, X)$ with $i=2,3,4$ are
\begin{align}
\label{eq:G23QT}
G_{2}^{\rm QT}(\varphi, X)&=\varphi^{D-2}\left[(D-1)h(\psi)-2\psi  h'(\psi)\right]\, \quad 
G_{3}^{\rm QT}(\varphi, X)=2\varphi^{D-3}h'(\psi)\, ,\\
\label{eq:G4QT}
G_{4}^{\rm QT}(\varphi, X)&=-\frac{1}{2}\varphi^{D-2}\psi^{(D-2)/2}\int \mathrm{d}\psi \psi^{-D/2}h'(\psi)\, ,
\end{align}
where, analogously to the Lovelock case, we have implicitly defined the \emph{characteristic function}
 \begin{equation}
 \label{eom_psiQT}
h(\psi)\equiv \psi + \sum_{n=2}^{\infty}\alpha_n {\color{black} \ell^{2n-2}}\frac{D-2n}{D-2}  \psi^n\, ,
\end{equation}
and where the primitive in Eq.~\eqref{eq:G4QT} must be chosen such that $G_{4}^{\rm QT}(1,1)=1/(D-2)$.\footnote{We fix \emph{en passant} a typo in equation (2.15) of~\cite{Bueno:2025gjg}, which should read as $G_{4}^{\rm QT}(1,0)=(1+n\alpha_n)/(D-2)$ or simply as $G_{4}^{\rm QT}(1,1)=1/(D-2)$.} Note the formal equivalence of these equations with~\eqref{eq:hornL},~\eqref{eq:G23L},~\eqref{eq:G4L} and~\eqref{eom_psiL}: for one to obtain the associated two-dimensional Horndeski theory of our QT gravities, one just needs to take the expression for the two-dimensional Horndeski theory~\eqref{eq:hornL} coming from a generic Lovelock gravity and perform the formal substitution $\lfloor D/2\rfloor \rightarrow \infty$. From this perspective, it becomes transparent how QT gravities generalize Lovelock gravities to those dimensions in which the latter theories cannot be simply defined. We close this subsection by noticing the following relation between the functions $G_{i}^{\rm QT}(\varphi, X)$:
\begin{equation}
G_{4,\varphi}^{\rm QT}=\frac{1}{2}G_{3}^{\rm QT}\, ,\quad G_{2,X}^{\rm QT}=-\frac{1}{2}G_{3,\varphi}^{\rm QT}\, ,
\end{equation}
where $G_{3,\varphi}^{\rm QT}$ denotes partial differentiation with respect to $\varphi$.

\subsection{Any two-dimensional Horndeski theory as the spherical reduction of a $D$-dimensional theory of gravity}

We have just seen that the spherical reduction of QT gravities gives rise to the two-dimensional Horndeski theory presented in Eq.~\eqref{eq:hornQT}. Nevertheless, as the functions $G_i^{\rm QT}(\varphi,X)$ are not completely general, it becomes clear that not all two-dimensional Horndeski theories come from the spherical reduction of  a QT gravity. Consequently, to identify a higher-dimensional theory of gravity for each two-dimensional Horndeski theory, we must drop some of the defining assumptions of QT gravities.

As it turns out, a successful strategy is provided by allowing for \emph{non-polynomial} curvature terms in the $D$-dimensional Lagrangian. Proceeding in this way, it was possible to obtain for the first time four-dimensional generalizations of QT gravities whose spherical reduction produces a two-dimensional Horndeski theory whose functions $G_i(\varphi,X)$ are given by Eqs.~\eqref{eq:G23QT} and~\eqref{eq:G4QT} with $D=4$~\cite{Bueno:2025zaj}. Allowing for more general non-polynomial theories, featuring covariant derivatives of the curvature, one may obtain by reduction on Eq.~\eqref{eq:sphermet} any two-dimensional Horndeski theory~\cite{Colleaux:2017ibe,Colleaux:2019ckh,Borissova:2026wmn,Borissova:2026krh}.

Without further ado, let us present a higher-dimensional (non-polynomial) origin for every two-dimensional Horndeski theory. To this aim, we will make use of the formalism developed in Ref.~\cite{Borissova:2026krh}. 
Define the following invariant with covariant derivatives of the curvature:
\begin{equation}
    \mathsf{\Omega}_4\equiv \nabla_e \mathsf{W}_{ab}{}^{cd}  \nabla^e \mathsf{W}_{cd}{}^{ab}\, .
\end{equation}
Consider now the following non-polynomial curvature invariants:
\begin{align}
\mathcal{I}_R&\equiv \frac{2\mathsf{R}}{D(D-1)}+ \frac{(D-2)(D-3) \mathsf{W}_3}{D^3-9D^2+26D-22} \left ( \frac{D-2}{\mathsf{W}_2}-\frac{2 \mathsf{Z}_3}{\mathsf{V}_5} \right)\,, \\
\mathcal{I}_\psi&\equiv\frac{\mathsf{R}}{D(D-1)}+ \frac{\mathsf{W}_3}{D^3-9D^2+26D-22} \left ( \frac{D-2}{\mathsf{W}_2}+\frac{2 (D-3) \mathsf{Z}_3}{\mathsf{V}_5} \right)\,, \\
\mathcal{I}_{\frac{\Box \varphi}{\varphi}}&\equiv-\frac{\mathsf{R}}{2(D-1)}+ \frac{(D-2)^2\mathsf{W}_3}{2(D^3-9D^2+26D-22)\mathsf{W}_2}  + \frac{(D-4) \mathcal{I}_\psi}{2}\,, \\
\notag 
\mathcal{I}_{(\mathcal{D}^2 \varphi)^2}&\equiv\frac{\mathsf{R}^2}{8(D-1)^2}- \frac{(D-2)^2 \mathsf{R}\mathsf{W}_3}{4(D-1)(D^3-9D^2+26D-22)\mathsf{W}_2}  + \frac{(D-2)^4 \mathsf{W}_3^2}{8(D^3-9D^2+26D-22)^2\mathsf{W}_2^2} \\&- \frac{(D^3-9D^2+26D-22) \mathsf{W}_2 \mathsf{Y}_3}{(D-1)(D-2)^3(D-3)\mathsf{W}_3}
+ \frac{D \mathsf{Z}_2}{2(D-1)(D-2)^2}\notag \\&+\frac{(D-4) \mathcal{I}_\psi}{2} \left(\mathcal{I}_{\frac{\Box \varphi}{\varphi}}-\frac{(D-4)\mathcal{I}_\psi}{4} \right) \,, \\
\mathcal{I}_{\frac{X}{\varphi^2}}&\equiv\frac{4 \mathsf{\Omega}_4\mathsf{W}_2- \nabla_e \mathsf{W}_2\nabla^e \mathsf{W}_2  }{8(D-1) \mathsf{W}_2^2}\,, \quad \mathcal{I}_{\varphi}^2\equiv \frac{1}{\mathcal{I}_\psi+\mathcal{I}_{\frac{X}{\varphi^2}}}\,,\quad \mathcal{I}_X\equiv\mathcal{I}_{\varphi}^2\, \mathcal{I}_{\frac{X}{\varphi^2}}\,,
\end{align}
where we have defined
\begin{equation}
    \mathsf{V}_5\equiv\mathsf{W}_2 \mathsf{Y}_3-\frac{(D-2)^2(D-3) \mathsf{Z}_2 \mathsf{W}_3}{D^3-9D^2+26D-22}\,.
\end{equation}
These invariants correspond to the $D$-dimensional generalization of those obtained in Ref.~\cite{Borissova:2026krh} for the $D=4$ case. We will be interested in performing the spherical reduction of $D$-dimensional theories of pure gravity $\mathcal{L}_{\rm D}\left (g^{ab}, \mathsf{R}_{abcd},\nabla_a\right )$ constructed from these non-polynomial curvature invariants featuring explicit covariant derivatives of the curvature. 

On top of Eq.~\eqref{eq:sphermet}, the previous curvature invariants evaluate to
\begin{align}
   \left.  \mathcal{I}_R \right \vert_\gamma &=R\,, \quad \left. \mathcal{I}_\psi \right \vert_\gamma=\frac{1-X}{\varphi^2}\,, \quad \left. \mathcal{I}_{\frac{\Box \varphi}{\varphi}} \right \vert_\gamma=\frac{\Box \varphi}{\varphi}\,, \quad \left. \mathcal{I}_{(\mathcal{D}^2 \varphi)^2} \right \vert_\gamma=\frac{1}{\varphi^2} \nabla_\mu \nabla_\nu \varphi \nabla^\mu \nabla^\nu \varphi\,, \\
   \left. \mathcal{I}_{\frac{X}{\varphi^2}}\right \vert_\gamma &=\frac{X}{\varphi^2}\,, \quad \left. \mathcal{I}_{\varphi}^2\right \vert_\gamma =\varphi^2\,, \quad \left. \mathcal{I}_X\right \vert_\gamma =X\,.
\end{align}
As a result, if we consider the higher-derivative theory
\begin{align}\notag
\mathcal{L}_D^{\rm NP}\left (g^{ab}, \mathsf{R}_{abcd},\nabla_a\right )&=\tilde{G}_2(\mathcal{I}_\varphi, \mathcal{I}_X)-\tilde{G}_3(\mathcal{I}_\varphi, \mathcal{I}_X)\mathcal{I}_{\frac{\Box \varphi}{\varphi}} +\tilde{G}_4(\mathcal{I}_\varphi, \mathcal{I}_X)\mathcal{I}_R  \\& \label{eq:Daction}  -2 \mathcal{I}_{\varphi}^2 \tilde{G}_{4,\mathcal{I}_X}(\mathcal{I}_\varphi, \mathcal{I}_X) \left\lbrace \mathcal{I}_{\frac{\Box \varphi}{\varphi}}^2-\mathcal{I}_{(\mathcal{D}^2 \varphi)^2}\right\rbrace \,,    
\end{align}
then the spherical reduction on top of~\eqref{eq:sphermet} yields precisely the two-dimensional theory \eqref{eq:2dgenaction} with $\mathcal{L}_{\rm 2d}$ as in \eqref{eq:horngen} with the identifications
\begin{equation}
    G_2(\varphi,X)= \varphi^{D-2} \tilde{G}_2(\varphi,X)\,,  \quad G_3(\varphi,X) =\varphi^{D-3} \tilde{G}_3(\varphi,X)\,, \quad G_4(\varphi,X)=\tilde{G}_4(\varphi,X)\varphi^{D-2} \,.
\end{equation}
Consequently, we prove that any two-dimensional Horndeski theory may be obtained via the spherical reduction of $D$-dimensional theories of pure gravity. Let us close by emphasizing the fact that this correspondence is not one-to-one---indeed, there are infinitely many higher-dimensional theories whose spherical reduction on Eq.~\eqref{eq:sphermet} provides the same two-dimensional Horndeski theory, as may be seen by adding to $\mathcal{L}_{\rm D}^{\rm NP}$ in Eq.~\eqref{eq:Daction} any combination of curvature invariants that vanish on top of~\eqref{eq:sphermet}.

\section{Low-temperature dynamics}\label{sec:low temperature dynamics}
In this section, we use the effective Horndeski description to derive the universal low-temperature thermodynamics of SS $D$-dimensional black holes. We evaluate the effective JT partition function and obtain the corresponding logarithmic quantum correction and breakdown temperature.

\subsection{Near-horizon reduction to JT gravity}
Let us start by considering a general extremal black-hole solution of a $D$-dimensional theory whose spherically symmetric sector is described by an effective two-dimensional Horndeski theory. Assuming a regular double horizon, the near-horizon geometry is AdS$_2$. In terms of the metric functions appearing in Eq.~\eqref{eqr}, extremality implies $f(r_0)=f'(r_0)=0$,  $f''(r_0)>0$, so that near the horizon
\begin{equation}
    N(r)=N(r_0)+\mathcal O(r-r_0)\, , \qquad f(r)=\frac{(r-r_0)^2}{L^2}+\mathcal O\left[(r-r_0)^3\right]\, ,
\end{equation}
where $L^2=2/f''(r_0)$ is the AdS$_2$ radius. Here we have assumed that the lapse function $N(r)$ is regular and nonvanishing at the horizon. These expressions, when plugged into Eq.~\eqref{eomf}, provide 
the following constraints\footnote{Observe that extremality also implies $X(r_0)=0$, simplifying greatly the algebra.} 
\begin{equation}\label{r0}
G_2(r_0,0)=0\,, \qquad
\frac{\partial_r G_{2}(r_0,0)}{2 \partial_r G_{4}(r_0,0)}=\frac{1}{L^2}\, ,
\end{equation}
where the latter is obtained after differentiation. These are to be understood as equations that fix the horizon radius and the AdS$_2$ radius as a function of the parameters of the Lagrangian and the charges --- see section \ref{examp}.

Our goal is then to evaluate the partition function of two-dimensional Horndeski gravity, accounting for quantum fluctuations. This contains the spherical fluctuations of the $D$-dimensional theories and hence the fluctuations in the AdS$_2$ throat length, which is the leading quantum contribution. 
We begin with Horndeski theory in two dimensions and in Euclidean signature\footnote{Notice the minus sign in the definition of the Euclidean action. In the case of GR, we have $Z=\int \mathcal Dg \,\e{-S_{\text E}}$, with $S_{\text E}=-1/(16\pi \GN)\int \sqrt{g} R$.}
\cite{Padilla:2012ze}
\begin{eqnarray}
S_{\rm 2d}&=&-\frac{(D-2)\Omega_{D-2}}{16\pi G_{\rm N}}\int {\rm d}^2 x \sqrt{|\gamma|}\mathcal{L}_{\rm 2d}\, ,
\end{eqnarray}
where $ \mathcal{L}_{\rm 2d}$ was defined in \req{eq:horngen}. Furthermore, for the variational principle to be well posed, we need to include a boundary action,
\begin{eqnarray}\label{S_total}
S_{\rm total}=S_{\rm 2d}-\frac{(D-2)\Omega_{D-2}}{16\pi G_{\rm N}}B\, ,\, \qquad 
B\equiv\int_{\Sigma}{\rm d}u\sqrt{|h_{uu}|}\left[ F_3+2G_4 K+4\Box^h\varphi F_{4,Y}\right]\, ,
\end{eqnarray}
where $\Sigma$ is the boundary curve with normal vector $n^\mu n_\mu=1$ and extrinsic curvature $K=\nabla_\mu n^\mu$,
\begin{equation}
F_l\equiv \int_0^{\varphi_n}G_l(\varphi,Y+ z^2){\rm d}z\, , \quad Y\equiv h^{uu}\dot{\varphi}^2\, ,
\end{equation} 
with $\varphi_n=n^\mu\partial_\mu\varphi$, and $\dot{\varphi}=\diff \varphi/\diff u$.

We will quantize Eq.~\eqref{S_total} in the near-extremal regime, in which the black hole horizon is $r_+=r_0+\delta r_+$, with $r_0$ the extremal horizon radius~\eqref{r0} and $\delta r_+\ll r_0$. As explained in Ref.~\cite{Iliesiu:2020qvm}, it is convenient to consider the separate contributions of the far and near-horizon regions. 

The far region, $r-r_0\gg\delta r_+$, is well approximated by
\begin{equation}
\diff s_{\text{far}}^2=N_0^2(r)f_0(r) \diff\tau^2+\frac{\diff r^2}{f_0(r)}\, ,
\end{equation}
where $f_0(r)$ is the metric function at zero temperature and $N_0(r)$ is the corresponding result of Eq.~\eqref{N(r)}. Since $f_0(r)$ has a double zero at $r=r_0$, this metric is smooth there only in the limit $\beta\to\infty$; for the large but finite $\beta$ of the actual near-extremal solution, extending it down to $r=r_0$ would produce a conical singularity. However, this does not pose a problem because the far-region approximation is only used away from $r_0$---and from the true horizon $r_+$---precisely where quantum fluctuations are strongly suppressed, so the would-be singular point is never actually reached. The far region's contribution to the partition function reduces to a constant shift set by the extremal mass of the black hole, together with a renormalization of the boundary term that renders the full action finite~\cite{Iliesiu:2020qvm}.  

The near-horizon region, on the other hand, is defined by $r-r_0\ll r_0$, with $r_0$ the horizon radius at extremality. In this limit the metric reads
\begin{equation}\label{BH_gauge}
\diff s_\text{nh}^2=N^2(r_0)f_\text{nh}(r) \diff\tau^2+\frac{\diff r^2}{f_\text{nh}(r)}\, , \quad f_\text{nh}(r)\approx \frac{(r-r_0)^2-\delta r_+^2}{L^2}\, ,
\end{equation}
where $L$ is the AdS$_2$ radius and $N(r_0)$, unlike the far-region lapse $N_0(r)$, is treated as constant across the throat, since it varies on the scale $r_0\gg r-r_0$. Note that $f_\text{nh}(r)$ vanishes at $r=r_0\pm\delta r_+$, the latter matching the horizon location $r_+$ of the far region. We also identify the dilaton with the radial coordinate (which we can always do due to diffeomorphism invariance) and write it in the form
\begin{equation}\label{varphi}
\varphi=\varphi_0(1+\phi)\, , \quad \phi\ll 1\, .
\end{equation}
We will consider a boundary separating the  far and near-horizon regions at constant radius $r_c$, satisfying $\delta r_+ \ll r_c-r_0\ll r_0$. Expanding the action to first order in $\phi$, and using the background equations of motion at $\phi=0$ to trade $G_2(\varphi_0,0)$ for $G_{4,\varphi}(\varphi_0,0)$, the result reads
\begin{eqnarray}\label{JT_bulk}
\mathcal{L}_{\rm 2d}&=&G_4R+\varphi_0 G_{4,\varphi}\left[R+\frac{2}{L^2}\right]\phi-\varphi_0G_3\Box\phi+\mathcal{O}(\phi^2)\, ,
\end{eqnarray}
recognizable as the standard Jackiw-Teitelboim structure, with $\phi(R+2/L^2)$ vanishing on the AdS$_2$ background. In the above expression all functions $G_i$ are evaluated at $\varphi=\varphi_0$ and $X=0$. It is also important to note that we assumed $X=\mathcal{O}(\phi^2)$, which we will later show to be consistent.

A few comments are in order. The zeroth order term is a topological invariant that equals the entropy of the extremal black hole. We require that $G_4\gg G_{\rm N}$ to avoid topology changes---more about this will be said later. 
Meanwhile, the box term is just a total derivative, and can be integrated by parts to give a boundary term. On the other hand, the equations of motion satisfied by the metric and dilaton are~\cite{Teitelboim:1983ux,Jackiw:1984je,Maldacena:2016upp},
\begin{align}
&R=-\frac{2}{L^2}\, ,\label{eq:RicAdS2}\\
&\nabla_\mu\nabla_\nu\phi-\nabla^2\phi g_{\mu\nu}+\frac{\phi}{L^2}g_{\mu\nu}=0\, .
\end{align}
The first equation implies that the classical geometry in the near-horizon region is locally AdS$_2$, in agreement with Eq.~\eqref{BH_gauge}. It is convenient to describe this geometry in Poincaré coordinates\footnote{For the coordinate transformation relating the black hole and Poincaré gauges, see Appendix~\ref{AdS_coordinates}.},
\begin{equation}
\diff s^2=\frac{L^2}{z^2}\left( \diff t^2+\diff z^2\right)\, .
\end{equation}
The general solution for the dilaton in this coordinate system is
\begin{equation}
\phi=\frac{A +Bt+C(t^2+z^2)}{z}\, .
\end{equation} 
Thus, near the AdS$_2$ boundary $X\sim\phi^2$, so it can be neglected from the perturbative expansion.

The conclusion that the bulk is AdS$_2$ is not restricted to the classical saddle. Since the dilaton appears linearly in the JT action, it acts as a Lagrange multiplier in the gravitational path integral, imposing the constraint~\eqref{eq:RicAdS2}. Consequently, pure JT gravity possesses no local bulk quantum degrees of freedom~\cite{Almheiri:2014cka,Maldacena:2016upp,Maldacena:2016hyu,Jensen:2016pah,Engelsoy:2016xyb}---see also Ref.~\cite{Sarosi:2017ykf} for a review. The remaining non-trivial dynamics is associated with the asymptotic boundary and is captured by a boundary reparametrization mode. 

\subsection{Boundary dynamics and the Schwarzian action}
To make this boundary degree of freedom explicit, we parametrize the boundary at $r=r_c$ by the curve
\begin{equation}\label{bdy_curve}
\mathcal{C}: X^a(u)=(t(u), z(u))\, , \qquad z(u)=\epsilon\, t'(u)\, ,
\end{equation}
as shown in Appendix~\ref{AdS_coordinates}, where $\epsilon$ is a cutoff and $u\in[0,\beta)$. We impose boundary conditions by fixing the value of the dilaton and the proper length of the boundary,
\begin{equation}\label{bdy_conditions}
\left.\phi\right|_{\mathcal C}
=\frac{\phi_r}{\epsilon},
\qquad
\ell_{\mathcal C}
\equiv
\int_{0}^{\beta}\diff u\,\sqrt{h_{uu}}
=\frac{\beta L}{\epsilon}.
\end{equation}
Crucially, fixing the proper length of the boundary rather than its induced metric allows for time reparametrizations $\tau(u)$, corresponding to large diffeomorphisms. These are explicitly broken by the boundary action, leaving only a residual gauge symmetry that we will discuss below. On the other hand, imposing a Dirichlet condition on the dilaton fixes its boundary value and therefore removes fluctuations normal to the boundary. Matching the proper length in the black hole coordinates to the boundary condition~\eqref{bdy_conditions} gives 
\begin{equation}\label{phi_r}
\phi_r=\frac{L^2}{N(\varphi_0)\varphi_0}\, .
\end{equation}
Since we are still working within the near-horizon region, we must have $\phi_r\ll\epsilon$. Using Eqs.~\eqref{phi_r} and~\eqref{cutoff}, this condition translates into $r_c-r_0\ll r_0$, which is consistent with our near-horizon approximation.

Expanding the boundary term $B$ in the total action~\eqref{S_total} to linear order in the dilaton, we obtain
\begin{equation}\label{eq:bdyaction}
B=2 G_4 \int_{\Sigma} \diff u \sqrt{h}K
+G_3 \varphi_0\int_\Sigma \diff u \sqrt{h}\,\phi_n
+2G_{4,\varphi}\varphi_0\int_\Sigma \diff u \sqrt{h}\,\phi K
+\mathcal{O}(\phi^2)\, .
\end{equation}
This boundary term ensures that the variational problem for the two-dimensional dilaton gravity theory~\eqref{JT_bulk} is well posed. In particular, the second term cancels the boundary contribution arising from the total-derivative term in the bulk action. 

To evaluate the extrinsic curvature of the boundary curve~\eqref{bdy_curve}, we decompose the inverse metric in the tangent–normal frame,
\begin{equation}
K=h^{ab}\nabla_bn_a=(T^aT^b+n^an^b)\nabla_bn_a=T^aT^b\nabla_bn_a\, ,
\end{equation}
where the unit tangent vector to $\mathcal{C}$ is
\begin{equation}
T^a=\frac{z(u)}{L\sqrt{t'^2+z'^2}}\,\dot X^a(u)=\frac{z(u)}{L\sqrt{t'^2+z'^2}}\big(t'(u),z'(u)\big)\, ,
\end{equation}
and the unit normal is $n^a=z(u)\big(z'(u),-t'(u)\big)/(L\sqrt{t'^2+z'^2})$. Evaluating the covariant derivative gives
\begin{equation}
K=\frac{t'\left(t'^2+z'^2+zz''\right)-zz't''}{L(t'^2+z'^2)^{3/2}}\, . \label{K_sch}
\end{equation}
Expanding in $\epsilon$ according to Eq.~\eqref{bdy_curve} makes the dependence of $K$ on the Schwarzian derivative of $t(u)$ manifest,
\begin{equation}
K=\frac{1}{L}\left(1+\epsilon^2\,\text{Sch}[t(u),u]\right)+\mathcal{O}(\epsilon^4)\, ,\qquad \text{Sch}[t(u),u]\equiv\left(\frac{t''}{t'}\right)'-\frac{1}{2}\left(\frac{t''}{t'}\right)^2\, .
\end{equation}
Substituting this result into the boundary action~\eqref{eq:bdyaction}, we are left with a purely Schwarzian contribution,
\begin{equation}\label{Sch_B}
B=\frac{2G_{4,\varphi} L^2}{N(r_0)}\int_0^\beta \diff u ~\text{Sch}[t(u),u]\, ,
\end{equation}
having canceled the accompanying divergence with a counterterm sourced by the boundary contribution of the far region,
\begin{equation}
B_{\rm ct}=-\frac{2G_{4,\varphi} \varphi_0}{L}    \int_{\Sigma}\diff u \sqrt{h} \phi\, .
\end{equation}

Collecting all the pieces, the full partition function organizes as
\begin{eqnarray}\label{Z_total}
Z=\e{-\beta M_0}\e{S_0}Z_{\rm JT}\, ,\quad Z_{\rm JT}= \int \frac{\diff \mu[t]}{\SLdR}\exp\left\{{\frac{2\pi C}{\beta}\int_0^{2\pi} \diff u\,\text{Sch}[t(u),u]}\right\}\, ,
\end{eqnarray}
with the Schwarzian coefficient being
\begin{eqnarray}\label{ZJT}
 \quad C\equiv\frac{(D-2)\Omega_{D-2}}{4\pi G_{\rm N}}\frac{G_{4,\varphi}(\varphi_0,0)^2}{N(r_0)G_{2,\varphi}(\varphi_0,0)}\, .
\end{eqnarray}
Despite having started from a generic higher-derivative dilaton gravity in the near-horizon throat rather than pure JT gravity, the boundary path integral collapses onto the same universal Schwarzian quantum mechanics familiar from Einstein gravity. This is a direct consequence of the symmetry-breaking pattern of the near-AdS$_2$ throat: any two-dimensional dilaton gravity with an AdS$_2$ vacuum possesses an asymptotic $\SLdR$ isometry that is spontaneously broken by the choice of boundary curve~\eqref{bdy_curve}, and the resulting soft mode $t(u)$ is universally characterized by the Schwarzian action, independently of which higher-curvature terms are present in the parent $D$-dimensional theory~\cite{Iliesiu:2020qvm}. All the model-dependence is instead confined to the overall coupling $C$. Here it is dressed by $G_{2,\varphi}$ and $G_{4,\varphi}$ 
evaluated on the near-horizon background, which capture the effect of arbitrary higher-derivative corrections to Einstein gravity into a single effective coefficient. In this sense, Eq.~\eqref{ZJT} makes manifest that the low-energy dynamics of near-extremal black holes in general higher-curvature gravities falls into the same universality class as in Einstein gravity, with the details of the particular theory entering only through $C$.

The first two factors, $\e{-\beta M_0}$ and $\e{S_0}$ in Eq.~\eqref{Z_total}, together give the zero-temperature partition function, with the exponent combining into $-\beta F_0$: the mass term comes from the far-region contribution, as anticipated, while the entropy arises from the Einstein-Hilbert action evaluated in the near-horizon geometry,
\begin{eqnarray}
S_0=G_4S_{\rm EH}= \frac{(D-2)\Omega_{D-2}}{4G_{\rm N}} G_4 \chi(M)\, ,
\end{eqnarray}
where we take $G_4\gg G_{\rm N}$,\footnote{In GR, $G_4$ is proportional to the area, as dictated by Eq.~\eqref{G_GR}. Hence, this condition means large black hole size. However, when higher-curvature corrections to GR are present, additional terms involving the horizon radius $r_0$ correct the area term. These contributions are resummed precisely in the function $G_4$.} so that higher-genus contributions are quickly suppressed and we can set $\chi=1$.\footnote{As we will discuss in Section~\ref{sec:nebht}, the suppression of higher-genus contributions eventually breaks down at sufficiently low temperatures, where contributions from different topologies can no longer be neglected.} This reproduces exactly the classical entropy obtained from Wald's formula \eqref{Wald_entropy}. The genuinely dynamical piece, $Z_{\rm JT}$, descends instead from the boundary term~\eqref{Sch_B}, after rescaling $u\rightarrow 2\pi u/\beta$. Let us now examine this contribution more closely.

\subsection{Schwarzian partition function}
We now turn to the quantization of the boundary reparametrization mode appearing in Eq.~\eqref{ZJT}. In particular, we explain the origin of the quotient by $\SLdR$ and the integration measure $\diff\mu[t]$. The Schwarzian action in Eq.~\eqref{ZJT} defines a theory of orientation-preserving reparametrizations of the circle, $\mathrm{Diff}({\rm S}^1)$. To make the circle manifest, we map from the Poincaré coordinate $t$ to the angular coordinate on the disk according to
\begin{equation}
t(u)=\tan\frac{\phi(u)}{2}\, , \qquad
\phi(u+2\pi)=\phi(u)+2\pi\, ,
\end{equation}
where $\phi(u)$ is a monotonically increasing function. The Schwarzian theory is moreover invariant under the right action of the M\"obius transformations
\begin{equation}
t\longrightarrow \frac{at+b}{ct+d}\, , \qquad
ad-bc=1\, .
\end{equation}
These transformations form the group $\SLdR$ and correspond to gauge redundancies of the boundary reparametrization mode. Consequently, the physical configuration space is
\begin{equation}
\frac{\mathrm{Diff}({\rm S}^1)}{\SLdR}\, ,
\end{equation}
and the corresponding gauge volume must be divided out in the path integral.

This space admits a natural symplectic structure.\footnote{More precisely,
$\mathrm{Diff}({\rm S}^1)/\SLdR$ can be regarded as a coadjoint orbit of the Virasoro group.
As such, it is equipped with a non-degenerate symplectic form, which allows for its
covariant quantization~\cite{Alekseev:1988ce}.} In terms of $\phi(u)$, the symplectic
form is~\cite{Witten:1987ty,Alekseev:1988ce}
\begin{equation}\label{symp}
\Omega=\int_0^{2\pi}\diff u\,
\left[
\frac{\diff\phi'\wedge\diff\phi''}{\phi'^2}
-\diff\phi\wedge(\diff\phi)'
\right]\, .
\end{equation}
The corresponding integration measure is given by the Pfaffian of the symplectic
form, $\diff\mu[\phi]={\rm Pf}(\Omega)\mathcal D\phi$, where ${\rm Pf}(\Omega)$
is the square root of the determinant of the antisymmetric operator associated with
$\Omega$. The Pfaffian can be represented as a functional integral over Grassmann
fields. To this end, we replace the exterior differential $\diff\phi$ by a
Grassmann-valued field $\psi$, so that the wedge products in Eq.~\eqref{symp} are
represented by products of Grassmann variables. 
The Schwarzian path integral can then be written as
\begin{equation}\label{super_theory}
Z_{\rm JT}=
\int\frac{\mathcal D\phi\,\mathcal D\psi}{\SLdR}
\exp\left\{
-\frac12\int_0^{2\pi}\diff u\,
\left[
\frac{\phi''^2}{g^2\phi'^2}
-\frac{\phi'^2}{g^2}
+\frac{\psi''\psi'}{\phi'^2}
-\psi'\psi
\right]
\right\}\, ,
\end{equation}
where, for later convenience, we have introduced
$g^2\equiv\beta/(2\pi C)$.

A key observation is that the Schwarzian is the generator of time translations, which acts through the Poisson brackets defined by the symplectic form \eqref{symp}. A more rigorous way of stating that is through the identity $\diff H=\iota_K \Omega$, where $K$ is the vector field that generates time translations\footnote{This would be $K=\phi' \frac{\diff }{\diff \phi}$.}. Hence, the full theory \eqref{super_theory}, which we might formally call $S_{\rm super}=H+\Omega$, is supersymmetric~\cite{Stanford:2017thb}. Explicitly, it is invariant under a transformation that swaps bosonic and fermionic degrees of freedom, which is generated by an equivariant differential form $Q=\diff- \iota_K$. It is straightforward to check that indeed $Q S_{\rm super}=0$ from the closedness of $\Omega$ and the defining property of the Hamiltonian $\diff H=\iota_K \Omega$. Then, we can add a supersymmetric-invariant term $\lambda QV$ to the action, with $V=g(K,\cdot)$ and $g$ any constant ($U(1)$-time translation invariant) metric. The resulting path integral would be independent of the parameter $\lambda$, but we can localize it by choosing $\lambda$ to be arbitrarily large. This means that the quantum fluctuations must satisfy $K=0$, and consequently $\diff H=0$, so the path integral is localized to the critical points of $H$. This is just a back-of-the-envelope explanation of what is known as the Duistermaat-Heckman (DH) formula~\cite{Duistermaat:1982vw}, which guarantees the one-loop exactness of the partition function under the only assumptions that the manifold is symplectic and the Lagrangian is the generator of a $U(1)$ symmetry. As explained above, in the case at hand these conditions are met, because the manifold is ${\rm Diff}(\rm S^1)/\SLdR$, which is symplectic, and the Schwarzian  is the Hamiltonian. 

Having established that the JT partition function is one-loop exact, let us evaluate it perturbatively~\cite{Saad:2019lba,Moitra:2021uiv}. We write $\phi(u)=u+g\epsilon(u)$, corresponding to a fluctuation about the saddle $\phi(u)=u$, and expand the action to quadratic order. We denote the resulting saddle-plus-one-loop expression by $Z_{\text{1-loop}}$, i.e.,
\begin{equation}\label{Z_part2}
Z_{\text{1-loop}}(g)=\int  \frac{\mathcal{D}(g\epsilon)\mathcal{D}\psi}{\SLdR}\exp{\left\{\frac{\pi}{g^2}\right\}}  \exp\left\{-\frac{1}{2}\int \diff u\,\left(\epsilon''^2-\epsilon'^2+\psi''\psi'-\psi'\psi\right)\right\}\, .
\end{equation}
Note that the bosonic integration variable in \eqref{Z_part2} is the perturbation $g\epsilon$. We fix the normalization of the bosonic measure by taking the Fourier coefficients of $g\epsilon$ to be integrated with the standard Lebesgue measure. Since $\epsilon(u)$ is periodic, we expand it as $\epsilon(u)=\sum_n\epsilon_n\e{i n u}$ and, after removing the modes $n=-1,0,1$ associated with the global $\SLdR$ transformations, we obtain
\begin{equation}\label{eq:measure_boson}
\frac{\mathcal{D}(g\epsilon)}{\SLdR}
\equiv\prod_{|n|\geq2}\,g\diff\epsilon_n
=\prod_{n\geq2}g^2\,\diff\epsilon_n\,\diff\epsilon_{-n}\,.
\end{equation}
Using the Fourier convolution identity\footnote{We use $\int \diff u\,f(u)g(u)\e{-\iu mu}=2\pi\sum_n f_n g_{m-n}$.} then gives
\begin{equation}\label{eq:fourier_boson}
\int\diff u\,({\epsilon''}^2-{\epsilon'}^2)
=2\pi\sum_{n\in\mathbb Z}(n^4-n^2)\epsilon_n\epsilon_{-n}.
\end{equation}
An analogous computation for the Grassmann field gives
\begin{equation}\label{changev_fermion}
\int\diff u\,(\psi''\psi'-\psi'\psi)
=2\pi\iu\sum_{n\in\mathbb Z}(n^3-n)\psi_n\psi_{-n}.
\end{equation}

Since the normalization of the bosonic measure has already been fixed by our choice of standard Lebesgue measure, we leave the overall normalization of the Grassmann functional measure unspecified. We denote it by a single constant $\mathcal \#$, into which we absorb every numerical factor generated by the computation below. We write
\begin{equation}\label{eq:Npsi}
\mathcal D\psi\equiv
\mathcal \#\prod_{n\geq2}\diff\psi_{n}\,\diff\psi_{-n}\,.  
\end{equation}

Combining expressions~\eqref{eq:measure_boson},
\eqref{eq:fourier_boson},~\eqref{changev_fermion} and
\eqref{eq:Npsi}, we write the saddle-plus-one-loop expression as
\begin{equation}
Z_{\text{1-loop}}=\exp\left\{\frac{\pi}{g^2}\right\}Z_\text{boson}Z_\text{fermion}\, ,\label{eq:Zoneloop}
\end{equation}
where
\begin{align}
Z_\text{boson}&\equiv\int \prod_{n\geq 2} (g^2 \diff \epsilon_n \diff \epsilon_{-n})\exp \left\{-2\pi\sum_{n \geq 2}(n^4-n^2)\epsilon_n \epsilon_{-n} \right\}\,,\\ 
Z_\text{fermion}&\equiv \mathcal \#\int \prod_{n\geq 2} ( \diff \psi_n \diff \psi_{-n})\exp \left\{-2\pi \iu \sum_{n \geq 2}(n^3-n)\psi_n \psi_{-n} \right\}\,.
\end{align}

Having factorized $Z_{\text{1-loop}}$ into bosonic and fermionic contributions, let us first evaluate the bosonic integral. Since $\epsilon(u)$ is real, the Fourier modes $\epsilon_n$ are not independent. They must satisfy $\epsilon_n ^\ast =\epsilon_{-n}$. Therefore, changing variables to the real and imaginary parts of $\epsilon$, i.e., $\epsilon_n =\epsilon_n ^{(\text{R})} + \iu \epsilon_n ^{(\text{I})}$, we get
\begin{align}
Z_\text{boson}
&= \int \prod_{n\geq 2} (g^2 \diff\epsilon_n ^{(\text{R})} \diff\epsilon_{n} ^{(\text{I})})\exp \left\{-2\pi\sum_{n \geq 2}(n^4-n^2)({\epsilon_n ^{(\text{R})} }^2 + {\epsilon_n ^{(\text{I})} }^2) \right\}\notag\\
&= \prod_{n\geq 2} g^2 \left(\int \diff\epsilon_n ^{(\text{R})} \exp \left\{-2\pi(n^4-n^2){\epsilon_n ^{(\text{R})} }^2  \right\}\right) ^2
= \prod_{n\geq 2} \frac{g^2}{2(n^4 - n^2)}\label{eq:Zboson}
\end{align}
On the other hand, following very similar steps to the ones explained above, we get for the fermionic factor
\begin{equation}\label{eq:Zfermion}
Z_\text{fermion}=\mathcal \#\prod_{n \geq 2} \left[-2\pi \iu(n^3-n)\right]\, .
\end{equation}
This expression contains a convention-dependent phase associated with the normalization of the Grassmann measure; as anticipated, we absorb it into $\mathcal \#$ and carry this single overall constant through the remainder of the computation. Combining the two factors~\eqref{eq:Zboson} and~\eqref{eq:Zfermion}, we obtain
\begin{equation}\label{Z1loop}
Z_{\text{1-loop}}=\#\exp\left\{\frac{\pi}{g^2}\right\}\prod_{n\geq 2}\frac{g^2}{n}\, .
\end{equation}
This is a divergent quantity, which we define using zeta-function regularization. The latter amounts to assigning a finite value to the divergent product through the analytic continuation of the associated zeta function. In the present case, the product in Eq.~\eqref{Z1loop} can be expressed in terms of the Riemann zeta function
\begin{equation}\label{eq:RiemannZeta}
\zeta(s)=\sum_{n=1}^{\infty} n^{-s} \, , \qquad{\rm Re}(s)>1.
\end{equation}
Using that $\zeta(0)=-1/2$ and $\zeta'(0)=-\frac{1}{2}\log{(2\pi)}$~\cite{Quine1993}, the Riemann zeta-regularized expression reads
\begin{equation}\label{eq:Zetareg}
\prod_{n\geq 2}\frac{g^2}{n}
=\e{[\zeta(0)-1]\log(g^2)+\zeta'(0)}
=\frac{1}{\sqrt{2\pi}}\frac{1}{g^3}\, .
\end{equation}
Substituting this result into Eq.~\eqref{Z1loop}, and recalling that the Schwarzian path integral is one-loop exact, we obtain the full JT partition function,
\begin{equation}\label{Z1loopreg}
Z_{\rm JT}=Z_{\text{1-loop}}=\#\frac{1}{g^3}\exp\left\{\frac{\pi}{g^2}\right\}\, ,\qquad g^2\equiv\frac{\beta}{2\pi C}\,,
\end{equation}
where, proceeding similarly as before, we absorbed the numerical factor---in this case, $1/\sqrt{2\pi}$---in Eq.~\eqref{eq:Zetareg} into $\#$. Notice that matching this result to the Schwarzian partition function of~\cite{Iliesiu:2020qvm} fixes $\#=(2\pi)^{-3/2}$.

\subsection{Near-extremal black hole thermodynamics}\label{sec:nebht}
Hence, we find that the partition function at small temperatures\footnote{Small compared to $1/r_0$. In the above derivation we relied on the presence of a large throat, such that there exists an intermediate regime in which $\delta r_+\ll r-r_0\ll r_0$.} is given by~\eqref{Z_total} together with~\eqref{Z1loopreg}, namely,
\begin{equation}\label{eq:Ztherm}
Z=\e{-\beta M_0+S_0}Z_{\rm JT}\, , \qquad \text{where}\qquad Z_{\rm JT}=\#\left(\frac{C}{\beta}\right)^{3/2}\exp\left\{\frac{2\pi^2 C}{\beta}\right\}\, .  
\end{equation}
As explained above, the first exponential contains the extremal mass $M_0$, which is a contribution from the far region geometry, together with the classical value of the entropy of the extremal black hole $S_0$. The latter is a topological contribution from the near-horizon region geometry. On the other hand, the following two terms come exclusively from the near-horizon region. However, they are of a different nature. While the third one is a classical contribution that corrects the extremal value of the free energy at low temperatures, the second one is quantum, due to the one-loop correction to the effective action. The entropy and the mean energy can be straightforwardly calculated\footnote{Note that the classical part of the low-temperature correction agrees with the one derived from \eqref{Wald_entropy}, expanding around the extremal horizon $r_+=r_0+\delta r_+$, with $\delta r_+=2\pi L^2 T/{\color{black} N(r_0)}$---see Eq.~\eqref{beta_rh}.}
\begin{eqnarray}\label{sfin}
S&=&\left(1-\beta \partial_\beta\right)\log Z=S_0+\frac{4\pi^2C}{\beta}+\frac{3}{2}\log{\frac{\#C}{\beta}}\, ,\\
U&=&-\partial_\beta\log Z= M_0+\frac{3}{2\beta}+\frac{2\pi^2 C}{\beta^2}\, .
\end{eqnarray}
Since $\log\#$ enters $\log Z$ only as an additive constant, it shifts the entropy and renders the precise value of $T_{\rm breakdown}$ normalization dependent~\cite{Iliesiu:2020qvm,Iliesiu:2022onk}. By contrast, it drops out of $U=-\partial_\beta\log Z$, so the corresponding energy scale is unambiguous. Interpreting $U=\langle M\rangle_\beta$ as the thermal expectation value of the black-hole energy and retaining only the classical Schwarzian saddle, we find
\begin{equation}\label{eq:massgap}
M_{\rm cl}(T)-M_0=2\pi^2CT^2+\mathcal O(T^3)
\equiv\frac{T^2}{M_{\rm gap}}+\mathcal O(T^3),
\qquad
M_{\rm gap}=\frac{1}{2\pi^2C}.
\end{equation}
The additional term $3T/2$ in $U-M_0$ arises from one-loop Schwarzian fluctuations. 

Before continuing, let us present some remarks on the coefficient $C$. First, notice that combining Eq.~\eqref{ZJT} with conditions~\eqref{r0} and~\eqref{Wald_entropy}, we can write 
the Schwarzian coefficient in terms of the Wald entropy as
\begin{equation}
C=\frac{L^2}{2\pi N(r_0)}
\left.\frac{\diff S}{\diff r_+}\right|_{r_+=r_0},
\label{eq:Centropy}
\end{equation}
where the derivative is evaluated along the near-extremal family with all conserved charges and gravitational couplings held fixed. This relation can also be derived from the first law of thermodynamics. On the other hand, $C$ controls the specific heat of the near-extremal black hole~\cite{Kolanowski:2024zrq}. At fixed conserved charges,
\begin{equation}\label{eq:CQ}
C_Q^{\text{cl}}\equiv T\left(\frac{\partial S_{\text{cl}}}{\partial T}\right)_Q
=4\pi^2 CT+\mathcal{O}(T^2)\,, \quad \text{so that }4\pi^2 C=\lim_{T\to0}\frac{C_Q^{\text{cl}}}{T}\,.
\end{equation}
Therefore, $C>0$ implies a positive specific heat near extremality and hence local thermodynamic stability in the fixed-charge ensemble, whereas $C<0$ signals an instability.

As the temperature is lowered within the genus-zero regime, the entropy becomes dominated by the quantum correction as
\begin{equation}\label{sfin2}
S \simeq S_0+\frac{3}{2}\log\frac{T}{ T_{\text{breakdown}}}\,,\qquad T_{\text{breakdown}}\sim C^{-1}\,,
\end{equation}
where $T_{\text{breakdown}}$ depends both on the particular details of the theory, via $C$, and on the choice of normalization, via $\#$.

There is a second, parametrically lower temperature scale at which the magnitude of the logarithmic correction becomes comparable to the extremal entropy itself. This is defined as~\cite{Saad:2019lba,Turiaci:2023wrh}
\begin{equation}\label{eq:Tnp}
-\frac{3}{2}\log\frac{T_{\rm np}}{T_{\text{breakdown}}}
\sim S_0\,,
\quad\Rightarrow\quad
T_{\rm np}\sim T_{\text{breakdown}}\exp\left\{-\frac{2S_0}{3}\right\}
\sim C^{-1}\exp\left\{-\frac{2S_0}{3}\right\}\,.
\end{equation}
Since at low temperatures the genus expansion is organized as~\cite{Mertens:2022irh}
\begin{equation}
Z\sim \sum_g (e^{S_0}(CT)^{3/2})^{1-2 g}\, ,
\end{equation}
this scale also marks the point at which higher-genus contributions become unsuppressed and the genus-zero approximation ceases to be reliable.

These results reproduce the universal near-extremal behavior established for Einstein-dilaton JT gravity and its higher-derivative generalizations~\cite{Iliesiu:2020qvm,Iliesiu:2022onk}. The $3/2\log T$ correction to the entropy, together with the dominance of the quantum term $3T/2$ over the classical $T^2$ correction to the energy at sufficiently low temperatures, match the low-temperature thermodynamics found for a wide class of near-extremal black holes. The coefficient $C$ also determines the leading specific heat and its sign. What makes this result useful in practice is that, within the genus-zero Schwarzian regime, however involved the parent $D$-dimensional Horndeski theory may be, its quantitative imprint on the temperature-dependent near-extremal thermodynamics is controlled through the single coefficient $C$ introduced in Eq.~\eqref{ZJT}. Once $C$ is evaluated for a given theory---through $G_{2,\varphi}$, $G_{4,\varphi}$, $N(r_0)$ and $G_{\rm N}$---the universal Schwarzian expressions for $S$ and $U$ immediately deliver its low-temperature thermodynamics. In this sense $C$ acts as a single dial along which the near-extremal behavior of different higher-curvature theories can be compared and computed. The range of validity of the genus-zero description is controlled separately by $S_0$, whose suppression of higher-genus contributions eventually breaks down at parametrically lower temperatures.

\section{Examples}\label{examp}
In this section we consider the quantum thermodynamics of particular near-extremal black hole geometries. Concretely, we focus on charged black holes in Lovelock gravity and regular black holes in Quasitopological gravity. As we saw earlier, these theories have second-order equations of motion around spherically symmetric backgrounds,  and therefore their dimensional reductions in the s-wave sector are particular instances of the most general two-dimensional Horndeski theory~\eqref{eq:horngen}. Therefore, the general analysis performed in the previous sections can be straightforwardly applied to these physically relevant cases.  

\subsection{Spherically symmetric extremal black holes in Lovelock gravity}\label{sec:ExtremeLove}

We now apply the general near-extremal analysis to spherically symmetric black holes in Lovelock gravity. Below we will distinguish between extremal solutions that arise in vacuum and those that are supported by a conserved charge. We write the characteristic polynomial as
\begin{equation}
h(\psi) = \psi + \sum_{n = 2}^K \alpha_n\ell^{2n-2}  \frac{D-2n}{D-2} \psi^n \,, \qquad K=\left\lfloor\frac{D-1}{2}\right\rfloor \, ,
\end{equation}
and we set $\alpha_1 = 1$. 

We shall consider Lovelock theory minimally coupled to a locally defined $(D-3)$-form potential $B$ with field strength $H = {\rm d} B$.  The Lorentzian action for this matter theory is
\begin{equation}
I_{\rm M}
=
-\frac{1}{2(D-2)!}
\int \dd^Dx\,\sqrt{|g|}\,
H_{\mu_1\cdots\mu_{D-2}}
H^{\mu_1\cdots\mu_{D-2}}\, .
\label{eq:dualMaxwell}
\end{equation}
The field strength obeys a Bianchi identity ${\rm d}H = 0$, which implies the existence of a conserved quantity,
\begin{equation}
Q = \frac{1}{\Omega_{D-2}} \int_{\rm S^{D-2}}  H \, .
\end{equation}
We should stress that our interest in this particular matter theory is purely pragmatic. We wish to study extremal black holes in Lovelock theory and, as we shall discuss below, for many choices of couplings the existence of spherically symmetric extremal black holes requires matter. Here we are not interested in understanding the contributions of this matter to the partition function, and hence we shall work in an ensemble where the charge of the field is fixed and therefore does not contribute additional fluctuating modes.

To study the matter theory, we expand $H$ in the complete basis of differential-form harmonics on $S^{D-2}$ and consistently truncate the massive Kaluza--Klein modes, leaving the only surviving component
\begin{equation}
H=\mathfrak{h}(t,r)\,\omega_{D-2}\, ,
\label{eq:Hansatz}
\end{equation}
where $\omega_{D-2}$ is the volume form of the unit $(D-2)$-sphere. The Bianchi identity immediately implies $\dd \mathfrak{h} = 0$ and hence 
\begin{equation}
\mathfrak{h}(t,r)=Q\, .
\label{Hflux}
\end{equation}
Substituting  Eq.~\eqref{Hflux} together with the metric ansatz \eqref{eq:sphermet} into the higher-dimensional action yields
\begin{equation}
I_{\rm M}
=
-\frac{\Omega_{D-2}}{2}
\int \diff ^2x\,\sqrt{|\gamma|}
\frac{Q^2}{\varphi^{D-2}} .
\label{eq:IM2d}
\end{equation}
Consequently, the two-dimensional theory governing the spherically symmetric perturbations of Lovelock gravity coupled to the dual gauge field is a Horndeski theory with
\begin{eqnarray}
G_2^{\rm LM}(\varphi,X)
&=&
G_2^{\rm L}(\varphi,X)
-
\frac{8\pi \GN Q^2}{(D-2)\varphi^{D-2}}\, ,
\\
G_3^{\rm LM}
&=&
G_3^{\rm L}\, ,
\\
G_4^{\rm LM}
&=&
G_4^{\rm L}\, .
\end{eqnarray}
After choosing the gauge $\varphi=r$ and \eqref{eqr}, the equations of motion imply that $N=1$, because $\partial_f\alpha =\partial_r \beta$, and 
\begin{equation}
h(\psi)=\frac{m}{r^{D-1}}-\frac{8\pi \GN Q^2}{(D-2)(D-3)r^{2(D-2)}}\, .
\end{equation}
Note that in GR $h(\psi)=\psi$, so this would reduce to the familiar Reissner--Nordstr\"om black hole in such a limiting case. 

Using the general expressions introduced earlier, we can write the thermodynamics explicitly for these charged Lovelock black holes
\begin{align}
    M &= \frac{(D-2) \Omega_{D-2}}{16 \pi G_{N}} r_+^{D-1} h(\psi_+) + \frac{\Omega_{D-2} Q^2}{2 (D-3) r_+^{D-3}} \, ,
    \\
    T &= \frac{1}{4 \pi r_+} \left[\frac{(D-1) r_+^2 h(\psi_+)}{h'(\psi_+)} - 2 - \frac{8 \pi \GN Q^2}{(D-2) h'(\psi_+) r_+^{2(D-3)}} \right] \,,
    \\
    S &= \frac{\Omega_{D-2}}{4 \GN} \sum_{n=1}^K n \alpha_n\ell^{2n-2} r_+^{D-2n} \, ,
\end{align}
where $\psi_+ = 1/r_+^2$. We will now examine the existence and properties of extremal Lovelock black holes, and assess the behavior of quantum corrections to their thermodynamics. Throughout we shall assume that the couplings are chosen such that $h'(x) \ge 0$. Not only does this ensure the absence of branch singularities, but this is a requirement to ensure the absence of ghosts when the theory involves a cosmological constant. This condition also ensures monotonicity of the entropy as a function of horizon radius. 

\paragraph{General extremality conditions.} It is convenient to define the function 
\begin{equation}
{\cal E}(\psi) = (D-1) h(\psi) - 2 \psi h'(\psi) \, .
\end{equation}
An extremal horizon of areal radius $r_0$ then requires that
\begin{equation}\label{eq:LL_ext_gen}
(D-2) {\cal E}(\psi_0) - 8 \pi \GN Q^2 \psi_0^{D-2} = 0 \, , \qquad \psi_0 = \frac{1}{r_0^2} \, . 
\end{equation}
The near-horizon geometry then takes the form of a metric on AdS$_2 \times$S$^{D-2}$ whose AdS$_2$ radius is
\begin{equation}
L^2 = \frac{h'(\psi_0)}{\Delta_0} \, , 
\end{equation}
with
\begin{equation}
\Delta_0 \equiv 2 \psi_0^2 h''(\psi_0) - (D-3) \psi_0 h'(\psi_0) + 8 \pi \GN Q^2 \psi_0^{D-2}\, .
\end{equation}
Therefore, for a regular AdS$_2$ throat we require in addition to $h'(\psi_0) > 0$ that $\Delta_0 > 0$.

The Horndeski functions at the extremal horizon are given by
\begin{equation}
G_{4,\varphi}^{\rm LM}(r_0,0)
=
r_0^{D-3}h'(\psi_0)\, ,
\qquad
G_{2,\varphi}^{\rm LM}(r_0,0)
=
\frac{2r_0^{D-3}h'(\psi_0)}{L^2}\, .
\end{equation}
It then follows that the Schwarzian coefficient introduced in Eq.~\eqref{ZJT} becomes
\begin{equation}
C_{\rm L}
=
\frac{(D-2)\Omega_{D-2}}{8\pi \GN}
L^2r_0^{D-3}h'(\psi_0) \, .
\end{equation}
Given our assumptions above, this quantity is positive.

\paragraph{Extremal Lovelock black holes in vacuum.} In vacuum, the charge-dependent terms in the above expressions vanish. If we assume that $\alpha_n > 0$ then extremal vacuum solutions are precluded in Lovelock theory. However, allowing for some $\alpha_n < 0$ it is possible to have such solutions while still satisfying our assumptions above. 

SS vacuum extremal black holes are absent in Einstein and Einstein–Gauss–Bonnet gravity. The first non-trivial examples in the class of theories considered here arise in $D=7$. Let us illustrate this explicitly with an example. Consider the Lovelock theory in $D = 7$ with 
\begin{equation}
h(\psi) = \psi - \ell^2 \psi^2 + \frac{2}{5} \ell^4 \psi^3,\, \qquad \alpha_{2} = - \frac{5}{3} \, , \quad \alpha_{3} = 2.
\end{equation}
It is easy to verify that $h'(\psi) > 0$ for all real $\psi$. Meanwhile,
\begin{equation}
{\cal E}(\psi) = 2 \psi(2 - \ell^2 \psi) \quad \Rightarrow  \quad r_0^2 = \frac{\ell^2}{2} \, .
\end{equation}
Hence at the extremal horizon we have
\begin{equation}
\Delta_0 = \frac{8}{\ell^2} \, , \quad L^2 = \frac{9}{40} \ell^2 \, ,
\end{equation}
and therefore all the physicality conditions are satisfied. In this case we also note that the Schwarzian coefficient is
\begin{equation}
C_{\rm L} = \frac{81 \pi^2}{1280} \frac{\ell^6}{\GN} \, ,
\end{equation}
while the extremal mass and entropy are
\begin{equation}
M_0 = \frac{3 \pi^2 \ell^4}{64 \GN} \, , \qquad S_0 = \frac{55 \pi^3 \ell^5}{48 \sqrt{2} \GN}\, . 
\end{equation}

More generally, a Lovelock theory with characteristic polynomial $h(x)$ will admit extremal black holes as vacuum solutions provided that
\begin{equation}
{\cal E}(\psi_0) = 0 \, , \qquad h'(\psi_0) > 0 \,, \qquad {\cal E}'(\psi_0) < 0 ,
\end{equation}
where the final condition is equivalent to $\Delta_0 > 0$ in vacuum, as $\Delta_0 = - \psi_0 {\cal E}'(\psi_0)$. Moreover,  the condition $h'(x) > 0$ ensures that the extremal entropy is positive. To see this, note that the extremal entropy can be rewritten in an integral form,
\begin{equation}
S_0 = \frac{(D-2) \Omega_{D-2}}{8 \GN} \int_{\psi_0}^\infty \dd x \, x^{-D/2} h'(x) \, .
\end{equation}
The integral is convergent in this form due to the $x^{-D/2}$ factor dominating polynomial contributions in $h'(x)$, which is guaranteed because $h(x)$ is a polynomial of at most order $\left\lfloor({D-1})/2\right\rfloor$. Hence the extremal entropy is an integral of a strictly positive integrand, and is therefore positive. 

Finally, let us emphasize that while vacuum extremal black holes are possible in Lovelock theory, the extremal horizon radius and AdS$_2$ length scale of the throat are proportional to the higher-curvature couplings. This means the higher-curvature terms will be of the same order as or larger than the Einstein ones, meaning that an effective field theory approach to the vacuum extremal black holes is not possible.

\paragraph{Extremal Lovelock black holes at fixed charge.} We now return to nonzero fixed charge $Q$. Solving Eq.~\eqref{eq:LL_ext_gen} for the charge gives
\begin{equation}
    Q^2 = \frac{(D-2)}{8 \pi \GN} \psi_0^{-(D-2)} {\cal E}(\psi_0) \, .
\end{equation}

Hence charge can support an extremal horizon at any prescribed $\psi_0 > 0$ whenever
\begin{equation}
{\cal E}(\psi_0) > 0 \, .
\end{equation}
This condition is far less restrictive than the vacuum condition ${\cal E}(\psi_0) = 0$.  In particular, charge supports extremal horizons in Einstein gravity and for broad regions of Lovelock parameter space where vacuum extremality is not possible. 

The simplest case is that of Reissner-Nordstr{\"o}m, for which we have $\alpha_n  =0$ for $n > 1$ giving
\begin{equation}
r_0^{2(D-3)} = \frac{8 \pi \GN Q^2}{(D-2)(D-3)}
\end{equation}
and
\begin{equation}
S_0 = \frac{\Omega_{D-2} r_0^{D-2}}{4 \GN} \,, \qquad L^2 = \frac{r_0^2}{(D-3)^2} \,, \qquad C_{\rm RN} = \frac{(D-2)\Omega_{D-2}}{8\pi \GN (D-3)^2} r_0^{D-1} \, .
\end{equation}

As a higher-curvature example, consider a charged Einstein--Gauss--Bonnet black hole in $D=5$, for which
\begin{equation}
h(\psi)=\psi+\frac{\alpha_2\ell^2}{3}\psi^2\,.   
\end{equation}
The extremality condition gives $r_0^4=4\pi G_{\rm N}Q^2/3$, as in the Reissner--Nordstr\"om case, while the extremal entropy, the AdS$_2$ radius, and the Schwarzian coefficient are
\begin{equation}
S_0=\frac{\pi^2}{2G_{\rm N}}
\left(r_0^3+2\alpha_2\ell^2r_0\right)\,,
\quad
L^2=\frac{1}{4}
\left(r_0^2+\frac{2\alpha_2\ell^2}{3}\right)\,,
\quad
C_{\rm GB}=
\frac{3\pi}{16G_{\rm N}}
\left(r_0^2+\frac{2\alpha_2\ell^2}{3}\right)^2\,.
\end{equation}
Consequently, from~\eqref{eq:massgap} we observe $
M_{\rm gap}
=8G_{\rm N}/
\left[3\pi^3\left(r_0^2+2\alpha_2\ell^2/3\right)^2\right]$, which agrees with the asymptotically flat limit of the result obtained in Ref.~\cite{Alvarado:2026kio}.

Contrary to the vacuum case, the presence of charge allows for the existence of large extremal black holes without resorting to large higher-curvature couplings. For fixed Lovelock couplings, the behavior of the large-charge regime is governed by the Einstein gravity terms, with the Lovelock ones providing subleading corrections,
\begin{equation}
r_0^{2(D-3)} = \frac{8 \pi \GN Q^2}{(D-2)(D-3)}\left[1 + \mathcal O\left(\alpha_n\frac{\ell^{2n-2}}{r_0^{2(n-1)}} \right) \right] \, .
\end{equation}
With charge, therefore, it is easy to realize $S_0 \gg 1$ without pushing the theory outside of an effective field theory regime.

\subsection{Pure-gravity regular black holes in Quasitopological gravity}

We now apply the preceding near-extremal analysis to vacuum regular black
holes in Quasitopological gravity. The theories and their spherical
reduction were reviewed above, so here we recall only the ingredients
needed for our analysis. In contrast to Lovelock gravity, the
characteristic function $h(\psi)$ of a Quasitopological theory can contain
terms of arbitrarily high order in $\psi$,
\begin{equation}
h(\psi)
=
\psi
+
\sum_{n=2}^{\infty}
\alpha_n\ell^{2n-2}
\frac{D-2n}{D-2}\,
\psi^n\, .
\label{eq:QT_characteristic}
\end{equation}
For suitable choices of the $\alpha_n$, the theory will admit asymptotically flat, regular black holes as vacuum
solutions~\cite{Bueno:2024dgm}.

For the static, spherically symmetric ansatz, the vacuum equations reduce to
\begin{equation}
N'=0\, ,
\qquad
\frac{\dd}{\dd r}\left[r^{D-1}h(\psi)\right]=0\, ,
\qquad
\psi=\frac{1-f(r)}{r^2}\, .
\end{equation}
After fixing the normalization of the time coordinate, we may set $N=1$,
and the metric function is determined algebraically by
\begin{equation}
h(\psi)=\frac{m}{r^{D-1}}\, .
\label{eq:QT_master}
\end{equation}
The integration constant $m$ is related to the ADM mass according to
\begin{equation}
M
=
\frac{(D-2)\Omega_{D-2}}{16\pi \GN}\,m\, .
\end{equation}

\paragraph{General thermodynamics and extremality.} At a spherical horizon, $\psi_+=1/r_+^2$, the mass and temperature are
\begin{align}
M
&=
\frac{(D-2)\Omega_{D-2}}{16\pi \GN}
r_+^{D-1}h(\psi_+)\, ,
\\
T
&=
\frac{1}{4\pi r_+}
\left[
\frac{(D-1)r_+^2h(\psi_+)}{h'(\psi_+)}-2
\right]\, .
\end{align}
The Wald entropy can be written as
\begin{equation}
S(r_+)
=
-\frac{(D-2)\Omega_{D-2}}{8\GN}
\int^{\psi_+}\dd x\,x^{-D/2}h'(x)
+S_{\rm sph}\, ,
\label{eq:QT_entropy_integral}
\end{equation}
where the additive constant $S_{\rm sph}$ will be discussed below. Once
the complete higher-dimensional action has been specified, including
terms which do not contribute to the spherical equations of motion,
$S_{\rm sph}$ is fixed. Differentiating~\eqref{eq:QT_entropy_integral} gives the exact relation
\begin{equation}
\frac{\dd S}{\dd r_+}
=
\frac{(D-2)\Omega_{D-2}}{4\GN}
r_+^{D-3}h'(\psi_+)\, .
\label{eq:QT_entropy_monotonicity}
\end{equation}
Thus, if $h'>0$, the entropy is strictly increasing as a function of horizon radius. However, contrary to the Lovelock case, positivity of $h'$ does not guarantee positivity of the entropy. This will depend on the dimension, the choice of $h$, and the additive constant $S_{\rm sph}$. 

As in the Lovelock case, let us define
\begin{equation}
{\cal E}(\psi)
=
(D-1)h(\psi)-2\psi h'(\psi)\, .
\end{equation}
A vacuum extremal horizon satisfies
\begin{equation}
{\cal E}(\psi_0)=0\, ,
\qquad
\psi_0=\frac{1}{r_0^2}\, .
\label{eq:QT_extremality}
\end{equation}
The near-horizon geometry is AdS$_2\times \rm S^{D-2}$, with
\begin{equation}
L^2
=
\frac{h'(\psi_0)}{\Delta_0}\, ,
\qquad
\Delta_0
\equiv
2\psi_0^2h''(\psi_0)
-(D-3)\psi_0h'(\psi_0)\, .
\label{eq:QT_L2}
\end{equation}
In vacuum one also has the useful identity $\Delta_0=-\psi_0{\cal E}'(\psi_0)$. Thus, a regular AdS$_2$ throat requires
\begin{equation}
h'(\psi_0)>0\, ,
\qquad
{\cal E}'(\psi_0)<0\, .
\end{equation}

At extremality,  the corresponding two-dimensional Horndeski functions obey
\begin{equation}
G_{4,\varphi}^{\rm QT}(r_0,0)
=
r_0^{D-3}h'(\psi_0)\, ,
\qquad
G_{2,\varphi}^{\rm QT}(r_0,0)
=
\frac{2r_0^{D-3}h'(\psi_0)}{L^2}\, .
\end{equation}
It follows that the Schwarzian coefficient~\eqref{ZJT} now is\footnote{Logarithmic corrections of a distinct origin to the entropy of static and spherically symmetric black holes have also been reported within certain non-polynomial QT theories \cite{Colleaux:2026hat}.}
\begin{equation}
C_{\rm QT}
=
\frac{(D-2)\Omega_{D-2}}{8\pi \GN}
L^2r_0^{D-3}h'(\psi_0) \, .
\label{eq:CQT_general}
\end{equation}
Therefore, $C_{\rm QT}>0$ whenever $h'(\psi_0)>0$ and $\Delta_0>0$.

\paragraph{Regular black holes from pure gravity.}
Eq.~\eqref{eq:QT_master} also provides a simple mechanism for
resolving the curvature singularity. Suppose that there is a finite
$\psi_{\rm c}>0$ such that
\begin{equation}
h(0)=0\, ,
\qquad
h'(0)=1\, ,
\qquad
h'(\psi)>0
\quad\text{for}\quad
0\leq\psi<\psi_{\rm c}\, ,
\qquad
\lim_{\psi\rightarrow\psi_{\rm c}^{-}}h(\psi)=+\infty\, .
\label{eq:QT_regularity}
\end{equation}
Then $h$ has a smooth inverse on the positive reals. As $r\rightarrow 0$, the right-hand side of
\eqref{eq:QT_master} diverges and hence
\begin{equation}
\psi(r)\longrightarrow\psi_{\rm c}\, ,
\qquad
f(r)=1-\psi_{\rm c}r^2+o(r^2)\, .
\end{equation}
The black hole consequently possesses a regular de Sitter core with
curvature scale $\ell_{\rm c}=\psi_{\rm c}^{-1/2}$. The explicit requirement that $h$ diverges at its radius of convergence is crucial to obtain regular spacetimes as solutions. 

Assuming the conditions above for the elimination of the curvature singularity, the existence of an extremal solution follows directly. Solving for the mass parameter as a function of horizon radius gives
\begin{equation}
m(r_+)
=
r_+^{D-1}h\left(\frac{1}{r_+^2}\right)\, ,
\label{eq:QT_mass_function}
\end{equation}
Under the conditions above, we have:
\begin{equation}
m(r_+)\longrightarrow+\infty
\quad\text{as}\quad
r_+\rightarrow\ell_{\rm c}^{+}
\quad\text{and} \quad m(r_+)\longrightarrow+\infty \quad \text{as} \quad
r_+\rightarrow+\infty\, .
\end{equation}
Therefore, $m(r_+)$ has at least one minimum. Moreover,
\begin{equation}
\frac{\dd m}{\dd r_+}
=
r_+^{D-2}{\cal E}(\psi_+)\, ,
\end{equation}
so every stationary point of the mass satisfies the extremality condition
\eqref{eq:QT_extremality}. At such a point,
\begin{equation}
\frac{\dd^2m}{\dd r_+^2}\bigg|_{r_0}
=
2r_0^{D-3}\Delta_0\, .
\end{equation}
Hence $\Delta_0>0$ is precisely the condition that the extremal
configuration be a local minimum of the mass. In the case in
which this minimum is unique, $m>m_0$ gives an inner and an outer
horizon, $m=m_0$ gives an extremal black hole, and $m<m_0$ gives a
regular horizonless geometry.

\paragraph{The Hayward resummation.}
For odd $D$, a simple regular black hole is obtained by choosing
\begin{equation}
\alpha_n=\frac{D-2}{D-2n}\, .
\end{equation}
The characteristic function then becomes
\begin{equation}
h(\psi)=\frac{\psi}{1-\ell^2\psi}\, ,
\end{equation}
and the corresponding metric is
\begin{equation}
f(r)
=
1-\frac{mr^2}{r^{D-1}+\ell^2m}\, .
\label{eq:QT_Hayward}
\end{equation}
This is the higher-dimensional Hayward regular black hole, obtained as
a vacuum solution of a purely gravitational theory
\cite{Hayward:2005gi,Bueno:2024dgm, Bueno:2025zaj}. It has
$\psi_{\rm c}=1/\ell^2$ and hence a de Sitter core with curvature
scale $\ell$.

The mass function has a unique minimum. Imposing extremality gives
\begin{equation}
\psi_0
=
\frac{D-3}{(D-1)\ell^2}\, ,
\qquad
r_0^2
=
\ell^2\frac{D-1}{D-3}\, .
\end{equation}
At the extremal horizon,
\begin{equation}
h'(\psi_0)
=
\frac{(D-1)^2}{4}\, ,
\qquad
\Delta_0
=
\frac{(D-1)(D-3)^2}{4\ell^2}\, ,
\end{equation}
and hence
\begin{equation}
L^2
=
\frac{\ell^2(D-1)}{(D-3)^2}\, ,
\end{equation}
while
\begin{equation}
C_{\rm H}
=
\frac{(D-2)(D-1)^2\Omega_{D-2}}
{32\pi \GN(D-3)}
r_0^{D-1}\, .
\label{eq:C_Hayward}
\end{equation}

For example, in $D=5$ one finds
\begin{equation}
r_0^2=2\ell^2\, ,
\qquad
L^2=\ell^2\, ,
\qquad
C_{\rm H}
=
\frac{6\pi\ell^4}{\GN}\, .
\end{equation}
The entropy, with the canonical choice $S_{\rm sph}=0$, is
\begin{equation}
\GN S_{\rm H}(r_+)
=
\frac{\pi^2r_+^3}{2}
+3\pi^2\ell^2r_+
-\frac{3\pi^2\ell^4r_+}{4(r_+^2-\ell^2)}
-
\frac{15\pi^2\ell^3}{4}
{\rm arctanh}\left(\frac{\ell}{r_+}\right)\, .
\end{equation}
At extremality this becomes
\begin{equation}
\GN S_0
=
\frac{\pi^2\ell^3}{4}
\left[
13\sqrt{2}
-
15{\rm arctanh}\left(\frac{1}{\sqrt{2}}\right)
\right]
>0\, .
\end{equation}
Since $h'>0$, Eq.~\eqref{eq:QT_entropy_monotonicity} proves that
the entropy is positive for all
$r_+\geq r_0$.

\paragraph{A regular solution with negative extremal entropy.}
Positivity of the entropy in the Hayward example is not universal. To
demonstrate this, consider in odd $D$ the dimensionless couplings
\begin{equation}
\alpha_{2k}=0\, ,
\qquad
\alpha_{2k+1}
=
\frac{D-2}{D-4k-2}
\frac{\Gamma\left(k+\frac{1}{2}\right)}
{\sqrt{\pi}\,\Gamma(k+1)}\, .
\end{equation}
The corresponding characteristic function resums to
\begin{equation}
h(\psi)
=
\frac{\psi}{\sqrt{1-\ell^4\psi^2}}\, ,
\label{eq:h_SR}
\end{equation}
and the algebraic field equation can be solved explicitly to give
\begin{equation}
f(r)
=
1-\frac{mr^2}
{\sqrt{r^{2(D-1)}+\ell^4m^2}}\, ,
\label{eq:QT_SR_metric}
\end{equation}
which is again regular.  Although the coefficients of the powers of $\psi$ in $h(\psi)$ are
positive, the dimensionless couplings $\alpha_{2k+1}$ above the
critical order are negative. These interactions can therefore make
negative contributions to the Wald entropy.

The extremality function is
\begin{equation}
{\cal E}(\psi)
=
\frac{\psi}
{\left(1-\ell^4\psi^2\right)^{3/2}}
\left[
(D-3)-(D-1)\ell^4\psi^2
\right]\, ,
\end{equation}
and the extremal horizon is therefore located at
\begin{equation}
\psi_0
=
\frac{1}{\ell^2}
\sqrt{\frac{D-3}{D-1}}\, ,
\qquad
r_0^2
=
\ell^2\sqrt{\frac{D-1}{D-3}}\, .
\label{eq:SR_extremal_radius}
\end{equation}
At this point,
\begin{equation}
h'(\psi_0)
=
\left(\frac{D-1}{2}\right)^{3/2}\, ,
\qquad
\Delta_0
=
2(D-3)\psi_0h'(\psi_0)>0\, .
\end{equation}
Consequently, the near-horizon geometry is
AdS$_2\times \rm S^{D-2}$, with
\begin{equation}
L^2
=
\frac{r_0^2}{2(D-3)}\, .
\label{eq:SR_AdS2_radius}
\end{equation}
The Schwarzian coefficient is
\begin{equation}
C_{\rm SR}
=
\frac{(D-2)\Omega_{D-2}}{16\pi \GN(D-3)}
\left(\frac{D-1}{2}\right)^{3/2}
r_0^{D-1}>0\, .
\label{eq:C_SR}
\end{equation}
Thus, this solution has both a regular AdS$_2$ throat and a positive
Schwarzian coefficient.

For odd $D$, with the choice $S_{\rm sph}=0$, the Wald
entropy can be written as
\begin{equation}
S_{\rm SR}(r_+)
=
\frac{\Omega_{D-2}r_+^{D-2}}{4\GN}
{}_2F_1\left(
\frac{3}{2},
-\frac{D-2}{4};
1-\frac{D-2}{4};
\frac{\ell^4}{r_+^4}
\right)\, .
\label{eq:S_SR}
\end{equation}
Despite the regularity of the geometry and the positivity of
$h'(\psi)$, this entropy need not be positive. Consider, for example, $D=5$. The extremal and near-horizon
parameters become
\begin{equation}
r_0^2=\sqrt{2}\ell^2\, ,
\qquad
L^2=\frac{r_0^2}{4}
=\frac{\ell^2}{2\sqrt{2}}\, ,
\qquad
C_{\rm SR}
=
\frac{3\pi\sqrt{2}}{4}
\frac{\ell^4}{\GN}\, .
\end{equation}
The extremal entropy is
\begin{align}
S_0
&=
\frac{\Omega_3r_0^3}{4\GN}
{}_2F_1\left(
\frac{3}{2},-\frac{3}{4};
\frac{1}{4};
\frac{1}{2}
\right) \simeq
-1.684842\,
\frac{\Omega_3r_0^3}{4\GN}
<0\, .
\label{eq:negative_QT_entropy}
\end{align}
This gives an explicit regular vacuum black hole satisfying
\begin{equation}
h'(\psi_0)>0\, ,
\qquad
\Delta_0>0\, ,
\qquad
C_{\rm SR}>0\, ,
\end{equation}
but with negative extremal Wald entropy.

Since the entropy is strictly increasing, it crosses zero only once. Numerically, this occurs at
\begin{equation}
r_S\simeq1.24298\,r_0\, .
\end{equation}
It follows that
\begin{equation}
S_{\rm SR}(r_+)<0
\quad\text{for}\quad
r_0\leq r_+<r_S\, ,
\qquad
S_{\rm SR}(r_+)>0
\quad\text{for}\quad
r_+>r_S\, .
\end{equation}
These ratios are independent of $\ell$. Therefore, adjusting the
higher-curvature scale $\ell$ does not eliminate the negative entropy
region.

\paragraph{Entropy shifts and topological terms.} That the naive Wald entropy can be negative for otherwise seemingly well-defined theories merits further discussion. To this end, let us revisit the additive constant $S_{\rm sph}$ appearing in the entropy~\eqref{eq:QT_entropy_integral}. At the level of the two-dimensional theory, the
addition
\begin{equation}
\Delta I_2
=
\frac{S_{\rm sph}}{4\pi}
\left[
\int_{\mathcal M}\dd^2x\,\sqrt{|g|}\,{\cal R}^{(2)}
+
2\int_{\partial\mathcal M}\dd s\,\sqrt{|h|}\,K
\right]
\label{eq:spherical_Euler_shift}
\end{equation}
shifts the black hole entropy by $S_{\rm sph}$ but does
not change $h(\psi)$, $M$, $T$, $r_0$, $L$, $\Delta_0$, or the
Schwarzian coefficient. In the Horndeski description, the shift
changes $G_4$ by a constant but leaves $G_{4,\varphi}$ unchanged. 

If the two-dimensional theory arises from the dimensional reduction of a higher-dimensional theory, then in even dimensions such a shift can arise from the topological Euler density. However, this is not the only possible origin for such a shift. Quasitopological theories are defined by their properties in spherical symmetry. As a result, one could add to the action terms which, on spherical symmetry,  are nonzero but yet dynamically trivial, leaving the dynamical properties of the Quasitopological theory under consideration unaffected. These terms, be they polynomial curvature invariants in $D = 2n$ or non-polynomial ones in any dimension, can source precisely this type of shift in the entropy. For example, using the covariant invariants introduced
above, which satisfy
\begin{equation}
\left.{\cal I}_R\right|_{\gamma}=R\, ,
\qquad
\left.{\cal I}_{\varphi}\right|_{\gamma}=\varphi\, ,
\end{equation}
one may add
\begin{equation}
\Delta I_{\rm sph}
=
\frac{\lambda\ell^{D-2}}{16\pi \GN}
\int\dd^Dx\,\sqrt{|g|}\,
\frac{{\cal I}_R}{{\cal I}_{\varphi}^{D-2}}\, .
\label{eq:nonpolynomial_spherical_term}
\end{equation}
The two-dimensional contribution is then proportional to the Euler
invariant~\eqref{eq:spherical_Euler_shift} above with 
\begin{equation}
S_{\rm sph} = \frac{\lambda\Omega_{D-2} \ell^{D-2}}{4 \GN} \, .
\end{equation}
Hence, a mechanism of this kind can be used to shift the Wald entropy by any constant amount. As the coefficient is proportional to the higher-curvature length scale, this does not affect the emergence of the area law in the Einstein gravity limit. 

A negative Wald entropy is difficult to reconcile with its
interpretation as a thermodynamic entropy. It does not, by itself,
indicate a local instability of the system.
In
particular, for an extremal black hole satisfying
$h'(\psi_0)>0$ and $\Delta_0>0$, the Schwarzian coefficient
$C_{\rm QT}$ remains positive independently of the sign of $S_0$. However, this conclusion does not establish the stability of non-spherical
fluctuations or of the complete higher-dimensional theory.

The more immediate difficulty concerns the role of topology in the JT
path integral. A connected orientable surface of genus $g$ with $n$
asymptotic boundaries has Euler characteristic
\begin{equation}
\chi_{g,n} = 2 - 2 g - n \, ,
\end{equation}
and enters the gravitational path integral with weight
\begin{equation}
\e{S_0 \chi_{g,n}} \, .
\end{equation}
At fixed $n$, adding one handle changes the Euler characteristic by $-2$. Hence these topological contributions are relatively suppressed by~\cite{Saad:2019lba}
\begin{equation}
\frac{\e{S_0\chi_{g+1,n}}}
     {\e{S_0\chi_{g,n}}}
=
\e{-2S_0}\, .
\end{equation}
Thus, when the extremal entropy $S_0 \gg 1$, higher-genus surfaces are exponentially
suppressed relative to the genus-zero contribution. If instead $S_0<0$, higher-genus contributions
are enhanced and there is no small parameter controlling an expansion
about genus zero. 

After adding a spherically topological term, the extremal entropy
becomes
\begin{equation}
S_0\longrightarrow S_0^{\rm eff}=S_0+ S_{\rm sph}\, ,
\end{equation}
and the relative cost of adding a handle is correspondingly
\begin{equation}
\e{-2S_0}
\longrightarrow
\e{-2(S_0+S_{\rm sph})}\, .
\end{equation}
A theory for which $S_0^{\rm eff}\gg 1$ therefore possesses the desired
controlled genus expansion. Note that this crucially depends on adding judiciously chosen terms to the action. Simply adjusting the integration constant in the Wald entropy would not modify the Euclidean on-shell
action and hence would not change the weighting of topological contributions.

Two theories related by such a shift have identical local equations and
classical black hole solutions within spherical symmetry, but different
absolute entropies, on-shell actions, and, therefore, assign different weights to different genus contributions. Moreover, if
the higher-dimensional density is only ``spherically topological'' rather
than genuinely topological, the theories can already differ in their
classical dynamics beyond spherical symmetry, and in their perturbations. The requirement of a controlled JT genus expansion may therefore
constrain couplings which are invisible to the local spherical
equations of motion.

\section{Conclusions}\label{conclu}

In this paper, we have shown that the universal quantum correction to the semiclassical thermodynamics of near-extremal black holes, arising from fluctuations of the AdS$_2$ near-horizon throat, extends to broad classes of higher-curvature gravity. Specifically, our analysis applies to SSS black holes in any theory whose SS sector admits an effective description in terms of two-dimensional Horndeski gravity, including infinite families of higher-curvature theories. Apart from the extremal data $M_0$ and $S_0$, all theory-dependent information entering the temperature-dependent genus-zero Schwarzian sector is encoded in the effective coupling given in Eq.~\eqref{ZJT}, while the resulting universal near-extremal partition function and thermodynamic quantities are presented in Eqs.~\eqref{eq:Ztherm}--\eqref{eq:Tnp}.

An interesting observation concerns the sign of the effective Schwarzian coupling. For the regular black holes considered here, the conditions $h'(\psi_0)>0$ and $\Delta_0>0$, which select a regular and physically admissible extremal branch, imply \(C_{\rm QT}>0\). From the near-horizon perspective, this is precisely the sign required for the boundary reparametrization mode to have a positive quadratic Euclidean action and for the near-extremal specific heat to be positive. Thus, the conditions ensuring the physical consistency of the classical black hole solution translate into a healthy, ghost-free Schwarzian sector in the low-energy quantum theory. Although this correspondence provides a nontrivial consistency check, positivity of \(C_{\rm QT}\) alone does not establish the unitarity of the complete higher-dimensional theory, whose nonspherical fluctuations and ultraviolet degrees of freedom are not captured by the Schwarzian description. It would furthermore be interesting to explore whether $C$ is also necessarily positive in the thermodynamics of more general integrable Horndeski models studied in~\cite{Borissova:2026rbi}. If not, positivity of $C$ would serve as a viability condition on the space of theories. 

A natural extension of our work would be to explore more general higher-curvature theories, for which the SS sector will generically involve higher-order derivatives. It would be interesting to determine whether the same universal $S\sim 3/2\log T$ correction persists, which would indicate that the near-extremal throat dynamics continues to be controlled by the symmetry breaking pattern of the AdS$_2$ near-horizon region.

It would also be interesting to extend our analysis beyond the SS sector. In Einstein-Maxwell theory, non-spherical fluctuations organize into towers of two-dimensional Kaluza-Klein fields, with the massive modes found not to modify the leading logarithmic dependence on temperature, although the massless rotational sector can yield additional ensemble- and scale-dependent contributions~\cite{Iliesiu:2020qvm,Kolanowski:2024zrq}. Whether an analogous decoupling persists in higher-curvature theories, or whether their enlarged fluctuation spectrum introduces additional nearly gapless modes, remains an open question.

In discussions concerning the ultimate fate of regular black holes and their viability as physical objects, Quasitopological gravities have emerged as the first natural framework---beyond more or less \emph{ad hoc} two-dimensional models---in which a broad range of previously inaccessible dynamical questions can be investigated. Of particular relevance is the evaporation of regular black holes~\cite{Hayward:2005gi,Hossenfelder:2009fc,Frolov:2014jva,Carballo-Rubio:2018pmi,Barcelo:2020mjw,Cadoni:2023tse,Barenboim:2024dko,Barenboim:2025ckx,Carballo-Rubio:2026gwg,Arrechea:2026rua,Easson:2026vte}. The quantum correction derived in our analysis is expected to modify the evaporation dynamics of regular black holes in Quasitopological theories as extremality is approached, analogously to the case of charged black holes in GR \cite{Brown:2024ajk}. We leave a detailed analysis of this question for future work. In particular, it will be important to determine whether the correction merely changes the rate at which extremality is approached or qualitatively modifies the endpoint of evaporation.



\section*{Acknowledgments}
We thank Roberto Emparan and Julio Oliva for useful conversations. GvdV thanks Valentín Benedetti and Mario Solís for the many stimulating and insightful discussions over the years. Moreover, he is especially grateful to Javier M. Magán for his guidance and the many lessons on quantum gravity during the early stages of his research career. PB and GvdV were supported by a Proyecto de Consolidación Investigadora (CNS 2023-143822) from Spain’s Ministry of Science, Innovation and Universities, and by the grant PID2022-136224NB-C22, funded by MCIN/AEI/10.13039/501100011033/FEDER, UE. PAC is supported by a Ramón y Cajal fellowship (RYC2023-044375-I) and by a Proyecto de Generación de Conocimiento (PID2024-155685NB-C22) from Spain’s Ministry of Science, Innovation and Universities, and by an “Attract RyC” grant (Agujeros negros, física fundamental y ondas gravitacionales) from the University of Murcia. RAH is supported by a Willmore Fellowship at Durham University. JM is supported by the European Union's Horizon Europe research and innovation programme under the Marie Sk\l odowska-Curie grant agreement
No.~101202710 (QUBITTO). ÁJM is supported by the European Union’s Horizon Europe research and innovation programme under the Marie Sk\l odowska-Curie grant agreement No.~101202045 (RoBHin-ETG).

\appendix
\section{AdS$_2$ in different coordinates}\label{AdS_coordinates}
The near-horizon geometry $r-r_0\ll r_0$ at low temperatures $\delta r_+ \ll r_0$ is given in Eq.~\eqref{BH_gauge}
with 
\begin{equation}\label{beta_rh}
\tau\sim \tau+\beta\, , \qquad \beta=\frac{2\pi L^2}{N(r_0)\delta r_+}\, .
\end{equation}
If we define
\begin{equation}
\theta=\frac{2\pi \tau}{\beta}\,,\qquad
r-r_0=\delta r_+ \coth{\sigma}\, ,
\end{equation}
with $\theta\in [0,2\pi)$ and $\sigma\in (0,\infty)$, then
\begin{equation}
\diff s^2=L^2 \frac{\diff \theta^2 +\diff \sigma^2}{\sinh^2{\sigma}}\, .
\end{equation}
This is still AdS$_2$ in global coordinates. To cover the Poincaré patch we consider the transformation
\begin{equation}\label{global_to_poincare}
t=\frac{\sin{\theta}}{\cosh{\sigma}+\cos{\theta}}\, , \qquad z=\frac{\sinh{\sigma}}{\cosh{\sigma}+\cos{\theta}}\, ,
\end{equation}
such that
\begin{equation}
\diff s^2=L^2 \frac{\diff t^2+ \diff z^2}{z^2}\, .
\end{equation}
A boundary at constant radial coordinate $\delta r_+\ll r_c-r_0\equiv \rho_c$ is at $\sigma\approx \frac{\delta r_+}{\rho_c}\ll 1$. Taking that limit in Eq.~\eqref{global_to_poincare} we see that
\begin{eqnarray}
t&\approx&\tan{\left(\frac{\pi\tau}{\beta}\right)}\,,\\
z&\approx&\left(\frac{L^2}{{\color{black}N(r_0)}\rho_c}\right)t'(\tau)\, .
\end{eqnarray}
Comparing with Eq.~\eqref{bdy_curve}, we have 
\begin{equation}\label{cutoff}
\epsilon=\frac{L^2}{{\color{black}N(r_0)}\rho_c}\, .
\end{equation}
\bibliographystyle{JHEP-2}
\bibliography{Gravities.bib}

\providecommand{\href}[2]{#2}\begingroup\raggedright\begin{thebibliography}{100}

\bibitem{Bekenstein:1972tm}
J.~D. Bekenstein, {\it {Black holes and the second law}},  {\em Lett. Nuovo
  Cim.} {\bf 4} (1972) 737--740.

\bibitem{Hawking:1975vcx}
S.~W. Hawking, {\it {Particle Creation by Black Holes}},  {\em Commun. Math.
  Phys.} {\bf 43} (1975) 199--220. [Erratum: Commun.Math.Phys. 46, 206 (1976)].

\bibitem{Bekenstein:1973ur}
J.~D. Bekenstein, {\it {Black holes and entropy}},  {\em Phys. Rev.} {\bf D7}
  (1973) 2333--2346.

\bibitem{Hawking:1976ra}
S.~W. Hawking, {\it {Breakdown of Predictability in Gravitational Collapse}},
  {\em Phys. Rev. D} {\bf 14} (1976) 2460--2473.

\bibitem{Bardeen:1973gs}
J.~M. Bardeen, B.~Carter and S.~W. Hawking, {\it {The Four laws of black hole
  mechanics}},  {\em Commun. Math. Phys.} {\bf 31} (1973) 161--170.

\bibitem{Almheiri:2020cfm}
A.~Almheiri, T.~Hartman, J.~Maldacena, E.~Shaghoulian and A.~Tajdini, {\it {The
  entropy of Hawking radiation}},  {\em Rev. Mod. Phys.} {\bf 93} (2021), no.~3
  035002 [\href{http://arXiv.org/abs/2006.06872}{{\tt 2006.06872}}].

\bibitem{Gibbons:1976ue}
G.~W. Gibbons and S.~W. Hawking, {\it {Action Integrals and Partition Functions
  in Quantum Gravity}},  {\em Phys. Rev.} {\bf D15} (1977) 2752--2756.

\bibitem{Gibbons:1976pt}
G.~W. Gibbons and M.~J. Perry, {\it {Black Holes and Thermal Green's
  Functions}},  {\em Proc. Roy. Soc. Lond. A} {\bf 358} (1978) 467--494.

\bibitem{Hawking:1982dh}
S.~W. Hawking and D.~N. Page, {\it {Thermodynamics of Black Holes in anti-De
  Sitter Space}},  {\em Commun. Math. Phys.} {\bf 87} (1983) 577.

\bibitem{York:1986it}
J.~W. York, Jr., {\it {Black hole thermodynamics and the Euclidean Einstein
  action}},  {\em Phys. Rev. D} {\bf 33} (1986) 2092--2099.

\bibitem{Hawking:1995fd}
S.~W. Hawking and G.~T. Horowitz, {\it {The Gravitational Hamiltonian, action,
  entropy and surface terms}},  {\em Class. Quant. Grav.} {\bf 13} (1996)
  1487--1498 [\href{http://arXiv.org/abs/gr-qc/9501014}{{\tt gr-qc/9501014}}].

\bibitem{Banerjee:2010qc}
S.~Banerjee, R.~K. Gupta and A.~Sen, {\it {Logarithmic Corrections to Extremal
  Black Hole Entropy from Quantum Entropy Function}},  {\em JHEP} {\bf 03}
  (2011) 147 [\href{http://arXiv.org/abs/1005.3044}{{\tt 1005.3044}}].

\bibitem{Banerjee:2011jp}
S.~Banerjee, R.~K. Gupta, I.~Mandal and A.~Sen, {\it {Logarithmic Corrections
  to N=4 and N=8 Black Hole Entropy: A One Loop Test of Quantum Gravity}},
  {\em JHEP} {\bf 11} (2011) 143 [\href{http://arXiv.org/abs/1106.0080}{{\tt
  1106.0080}}].

\bibitem{Sen:2012cj}
A.~Sen, {\it {Logarithmic Corrections to Rotating Extremal Black Hole Entropy
  in Four and Five Dimensions}},  {\em Gen. Rel. Grav.} {\bf 44} (2012)
  1947--1991 [\href{http://arXiv.org/abs/1109.3706}{{\tt 1109.3706}}].

\bibitem{Ghosh:2019rcj}
A.~Ghosh, H.~Maxfield and G.~J. Turiaci, {\it {A universal Schwarzian sector in
  two-dimensional conformal field theories}},  {\em JHEP} {\bf 05} (2020) 104
  [\href{http://arXiv.org/abs/1912.07654}{{\tt 1912.07654}}].

\bibitem{Heydeman:2020hhw}
M.~Heydeman, L.~V. Iliesiu, G.~J. Turiaci and W.~Zhao, {\it {The statistical
  mechanics of near-BPS black holes}},  {\em J. Phys. A} {\bf 55} (2022), no.~1
  014004 [\href{http://arXiv.org/abs/2011.01953}{{\tt 2011.01953}}].

\bibitem{Iliesiu:2020qvm}
L.~V. Iliesiu and G.~J. Turiaci, {\it {The statistical mechanics of
  near-extremal black holes}},  {\em JHEP} {\bf 05} (2021) 145
  [\href{http://arXiv.org/abs/2003.02860}{{\tt 2003.02860}}].

\bibitem{Iliesiu:2022onk}
L.~V. Iliesiu, S.~Murthy and G.~J. Turiaci, {\it {Revisiting the logarithmic
  corrections to the black hole entropy}},  {\em JHEP} {\bf 07} (2025) 058
  [\href{http://arXiv.org/abs/2209.13608}{{\tt 2209.13608}}].

\bibitem{Boruch:2022tno}
J.~Boruch, M.~T. Heydeman, L.~V. Iliesiu and G.~J. Turiaci, {\it {BPS and
  near-BPS black holes in AdS$_{5}$ and their spectrum in $ \mathcal{N} $ = 4
  SYM}},  {\em JHEP} {\bf 07} (2025) 220
  [\href{http://arXiv.org/abs/2203.01331}{{\tt 2203.01331}}].

\bibitem{Kolanowski:2024zrq}
M.~Kolanowski, D.~Marolf, I.~Rakic, M.~Rangamani and G.~J. Turiaci, {\it
  {Looking at extremal black holes from very far away}},  {\em JHEP} {\bf 04}
  (2025) 020 [\href{http://arXiv.org/abs/2409.16248}{{\tt 2409.16248}}].

\bibitem{Penington:2019npb}
G.~Penington, {\it {Entanglement Wedge Reconstruction and the Information
  Paradox}},  {\em JHEP} {\bf 09} (2020) 002
  [\href{http://arXiv.org/abs/1905.08255}{{\tt 1905.08255}}].

\bibitem{Penington:2019kki}
G.~Penington, S.~H. Shenker, D.~Stanford and Z.~Yang, {\it {Replica wormholes
  and the black hole interior}},  {\em JHEP} {\bf 03} (2022) 205
  [\href{http://arXiv.org/abs/1911.11977}{{\tt 1911.11977}}].

\bibitem{Almheiri:2019qdq}
A.~Almheiri, T.~Hartman, J.~Maldacena, E.~Shaghoulian and A.~Tajdini, {\it
  {Replica Wormholes and the Entropy of Hawking Radiation}},  {\em JHEP} {\bf
  05} (2020) 013 [\href{http://arXiv.org/abs/1911.12333}{{\tt 1911.12333}}].

\bibitem{Balasubramanian:2022gmo}
V.~Balasubramanian, A.~Lawrence, J.~M. Magan and M.~Sasieta, {\it {Microscopic
  Origin of the Entropy of Black Holes in General Relativity}},  {\em Phys.
  Rev. X} {\bf 14} (2024), no.~1 011024
  [\href{http://arXiv.org/abs/2212.02447}{{\tt 2212.02447}}].

\bibitem{Climent:2024trz}
A.~Climent, R.~Emparan, J.~M. Magan, M.~Sasieta and A.~Vilar~L{\'o}pez, {\it
  {Universal construction of black hole microstates}},  {\em Phys. Rev. D} {\bf
  109} (2024), no.~8 086024 [\href{http://arXiv.org/abs/2401.08775}{{\tt
  2401.08775}}].

\bibitem{Preskill:1991tb}
J.~Preskill, P.~Schwarz, A.~D. Shapere, S.~Trivedi and F.~Wilczek, {\it
  {Limitations on the statistical description of black holes}},  {\em Mod.
  Phys. Lett. A} {\bf 6} (1991) 2353--2362.

\bibitem{Maldacena:1998uz}
J.~M. Maldacena, J.~Michelson and A.~Strominger, {\it {Anti-de Sitter
  fragmentation}},  {\em JHEP} {\bf 02} (1999) 011
  [\href{http://arXiv.org/abs/hep-th/9812073}{{\tt hep-th/9812073}}].

\bibitem{Nayak:2018qej}
P.~Nayak, A.~Shukla, R.~M. Soni, S.~P. Trivedi and V.~Vishal, {\it {On the
  Dynamics of Near-Extremal Black Holes}},  {\em JHEP} {\bf 09} (2018) 048
  [\href{http://arXiv.org/abs/1802.09547}{{\tt 1802.09547}}].

\bibitem{Maxfield:2020ale}
H.~Maxfield and G.~J. Turiaci, {\it {The path integral of 3D gravity near
  extremality; or, JT gravity with defects as a matrix integral}},  {\em JHEP}
  {\bf 01} (2021) 118 [\href{http://arXiv.org/abs/2006.11317}{{\tt
  2006.11317}}].

\bibitem{Maldacena:2019cbz}
J.~Maldacena, G.~J. Turiaci and Z.~Yang, {\it {Two dimensional Nearly de Sitter
  gravity}},  {\em JHEP} {\bf 01} (2021) 139
  [\href{http://arXiv.org/abs/1904.01911}{{\tt 1904.01911}}].

\bibitem{Maulik:2025phe}
S.~Maulik, A.~Mitra, D.~Mukherjee and A.~Ray, {\it {Logarithmic corrections to
  near-extremal entropy of charged de Sitter black holes}},  {\em JHEP} {\bf
  01} (2026) 156 [\href{http://arXiv.org/abs/2503.08617}{{\tt 2503.08617}}].

\bibitem{Blacker:2025zca}
M.~J. Blacker, A.~Castro, W.~Sybesma and C.~Toldo, {\it {Quantum corrections to
  the path integral of near extremal de Sitter black holes}},  {\em JHEP} {\bf
  08} (2025) 120 [\href{http://arXiv.org/abs/2503.14623}{{\tt 2503.14623}}].

\bibitem{Kapec:2023ruw}
D.~Kapec, A.~Sheta, A.~Strominger and C.~Toldo, {\it {Logarithmic Corrections
  to Kerr Thermodynamics}},  {\em Phys. Rev. Lett.} {\bf 133} (2024), no.~2
  021601 [\href{http://arXiv.org/abs/2310.00848}{{\tt 2310.00848}}].

\bibitem{Rakic:2023vhv}
I.~Rakic, M.~Rangamani and G.~J. Turiaci, {\it {Thermodynamics of the
  near-extremal Kerr spacetime}},  {\em JHEP} {\bf 06} (2024) 011
  [\href{http://arXiv.org/abs/2310.04532}{{\tt 2310.04532}}].

\bibitem{Maulik:2024dwq}
S.~Maulik, L.~A. Pando~Zayas, A.~Ray and J.~Zhang, {\it {Universality in
  logarithmic temperature corrections to near-extremal rotating black hole
  thermodynamics in various dimensions}},  {\em JHEP} {\bf 06} (2024) 034
  [\href{http://arXiv.org/abs/2401.16507}{{\tt 2401.16507}}].

\bibitem{Kapec:2024zdj}
D.~Kapec, Y.~T.~A. Law and C.~Toldo, {\it {Quasinormal corrections to
  near-extremal black hole thermodynamics}},  {\em JHEP} {\bf 06} (2025) 069
  [\href{http://arXiv.org/abs/2409.14928}{{\tt 2409.14928}}].

\bibitem{Modak:2025gvp}
A.~Modak, A.~Singh and B.~Panda, {\it {Logarithmic corrections for
  near-extremal Kerr-Newman Black holes}},  {\em JHEP} {\bf 03} (2026) 151
  [\href{http://arXiv.org/abs/2502.18173}{{\tt 2502.18173}}].

\bibitem{PandoZayas:2026vbg}
L.~A. Pando~Zayas and J.~Zhang, {\it {A Universality Theorem for the Quantum
  Thermodynamics of Near-Extremal Black Holes}},
  \href{http://arXiv.org/abs/2602.16767}{{\tt 2602.16767}}.

\bibitem{Banerjee:2021vjy}
N.~Banerjee, T.~Mandal, A.~Rudra and M.~Saha, {\it {Equivalence of JT gravity
  and near-extremal black hole dynamics in higher derivative theory}},  {\em
  JHEP} {\bf 01} (2022) 124 [\href{http://arXiv.org/abs/2110.04272}{{\tt
  2110.04272}}].

\bibitem{Rathi:2021aaw}
H.~Rathi and D.~Roychowdhury, {\it {Holographic JT gravity with quartic
  couplings}},  {\em JHEP} {\bf 10} (2021) 209
  [\href{http://arXiv.org/abs/2107.11632}{{\tt 2107.11632}}].

\bibitem{Alvarado:2026kio}
A.~Alvarado, A.~Anabalon, M.~Chernicoff, J.~Oliva, M.~Oyarzo, G.~Ortega and
  J.~Urbina, {\it {Logarithmic corrections to the entropy of near-extremal
  black holes in Einstein-Gauss-Bonnet}},
  \href{http://arXiv.org/abs/2603.24939}{{\tt 2603.24939}}.

\bibitem{Acito:2026mmf}
L.~Acito, M.~Chernicoff, J.~Oliva, C.~R. d.~A. Torres and M.~Sempe, {\it
  {Logarithmic corrections to the entropy of near-extremal black holes in New
  Massive Gravity}},  \href{http://arXiv.org/abs/2606.13546}{{\tt 2606.13546}}.

\bibitem{Despontin:2026xzg}
E.~Despontin, S.~Detournay, R.~Mancilla and C.~Toldo, {\it {Quantum corrections
  to the near-extremal thermodynamics of (warped) BTZ black holes}},
  \href{http://arXiv.org/abs/2607.08482}{{\tt 2607.08482}}.

\bibitem{Maldacena:2016upp}
J.~Maldacena, D.~Stanford and Z.~Yang, {\it {Conformal symmetry and its
  breaking in two dimensional Nearly Anti-de-Sitter space}},  {\em PTEP} {\bf
  2016} (2016), no.~12 12C104 [\href{http://arXiv.org/abs/1606.01857}{{\tt
  1606.01857}}].

\bibitem{Stanford:2017thb}
D.~Stanford and E.~Witten, {\it {Fermionic Localization of the Schwarzian
  Theory}},  {\em JHEP} {\bf 10} (2017) 008
  [\href{http://arXiv.org/abs/1703.04612}{{\tt 1703.04612}}].

\bibitem{Saad:2019lba}
P.~Saad, S.~H. Shenker and D.~Stanford, {\it {JT gravity as a matrix
  integral}},  \href{http://arXiv.org/abs/1903.11115}{{\tt 1903.11115}}.

\bibitem{Jackiw:1984je}
R.~Jackiw, {\it {Lower Dimensional Gravity}},  {\em Nucl. Phys. B} {\bf 252}
  (1985) 343--356.

\bibitem{Teitelboim:1983ux}
C.~Teitelboim, {\it {Gravitation and Hamiltonian Structure in Two Space-Time
  Dimensions}},  {\em Phys. Lett. B} {\bf 126} (1983) 41--45.

\bibitem{Gaikwad:2018dfc}
A.~Gaikwad, L.~K. Joshi, G.~Mandal and S.~R. Wadia, {\it {Holographic dual to
  charged SYK from 3D Gravity and Chern-Simons}},  {\em JHEP} {\bf 02} (2020)
  033 [\href{http://arXiv.org/abs/1802.07746}{{\tt 1802.07746}}].

\bibitem{Moitra:2019bub}
U.~Moitra, S.~K. Sake, S.~P. Trivedi and V.~Vishal, {\it {Jackiw-Teitelboim
  Gravity and Rotating Black Holes}},  {\em JHEP} {\bf 11} (2019) 047
  [\href{http://arXiv.org/abs/1905.10378}{{\tt 1905.10378}}].

\bibitem{Sachdev:2019bjn}
S.~Sachdev, {\it {Universal low temperature theory of charged black holes with
  AdS$_2$ horizons}},  {\em J. Math. Phys.} {\bf 60} (2019), no.~5 052303
  [\href{http://arXiv.org/abs/1902.04078}{{\tt 1902.04078}}].

\bibitem{Mertens:2022irh}
T.~G. Mertens and G.~J. Turiaci, {\it {Solvable models of quantum black holes:
  a review on Jackiw{\textendash}Teitelboim gravity}},  {\em Living Rev. Rel.}
  {\bf 26} (2023), no.~1 4 [\href{http://arXiv.org/abs/2210.10846}{{\tt
  2210.10846}}].

\bibitem{Turiaci:2023wrh}
G.~J. Turiaci, {\it {New insights on near-extremal black holes}},
  \href{http://arXiv.org/abs/2307.10423}{{\tt 2307.10423}}.

\bibitem{Martinez:1996uv}
C.~Martinez and J.~Zanelli, {\it {Back reaction of a conformal field on a
  three-dimensional black hole}},  {\em Phys. Rev. D} {\bf 55} (1997)
  3642--3646 [\href{http://arXiv.org/abs/gr-qc/9610050}{{\tt gr-qc/9610050}}].

\bibitem{Casals:2016odj}
M.~Casals, A.~Fabbri, C.~Mart{\'\i}nez and J.~Zanelli, {\it {Quantum
  Backreaction on Three-Dimensional Black Holes and Naked Singularities}},
  {\em Phys. Rev. Lett.} {\bf 118} (2017), no.~13 131102
  [\href{http://arXiv.org/abs/1608.05366}{{\tt 1608.05366}}].

\bibitem{Emparan:2020znc}
R.~Emparan, A.~M. Frassino and B.~Way, {\it {Quantum BTZ black hole}},  {\em
  JHEP} {\bf 11} (2020) 137 [\href{http://arXiv.org/abs/2007.15999}{{\tt
  2007.15999}}].

\bibitem{Battista:2023iyu}
E.~Battista, {\it {Quantum Schwarzschild geometry in effective field theory
  models of gravity}},  {\em Phys. Rev. D} {\bf 109} (2024), no.~2 026004
  [\href{http://arXiv.org/abs/2312.00450}{{\tt 2312.00450}}].

\bibitem{Chernicoff:2024dll}
M.~Chernicoff, G.~Giribet, J.~Moreno, J.~Oliva, R.~Rojas and C.~R. d.~A.
  Torres, {\it {Quantum backreactions in (A)dS3 massive gravity and logarithmic
  asymptotic behavior}},  {\em Phys. Rev. D} {\bf 110} (2024), no.~4 044021
  [\href{http://arXiv.org/abs/2404.10127}{{\tt 2404.10127}}].

\bibitem{Wang:2025fmz}
Z.-L. Wang and E.~Battista, {\it {Dynamical features and shadows of quantum
  Schwarzschild black hole in effective field theories of gravity}},  {\em Eur.
  Phys. J. C} {\bf 85} (2025), no.~3 304
  [\href{http://arXiv.org/abs/2501.14516}{{\tt 2501.14516}}].

\bibitem{Frassino:2024bjg}
A.~M. Frassino, R.~A. Hennigar, J.~F. Pedraza and A.~Svesko, {\it {Quantum
  Inequalities for Quantum Black Holes}},  {\em Phys. Rev. Lett.} {\bf 133}
  (2024), no.~18 181501 [\href{http://arXiv.org/abs/2406.17860}{{\tt
  2406.17860}}].

\bibitem{Climent:2024wol}
A.~Climent, R.~A. Hennigar, E.~Panella and A.~Svesko, {\it {Nucleation of
  charged quantum de-Sitter$_{3}$ black holes}},  {\em JHEP} {\bf 05} (2025)
  086 [\href{http://arXiv.org/abs/2410.02375}{{\tt 2410.02375}}].

\bibitem{Mendez-Zavaleta:2026rgg}
J.~A. M{\'e}ndez-Zavaleta, E.~Rojas and J.~J. Su{\'a}rez-Garibay, {\it
  {Classical emergence of the quantum-backreacted BTZ black hole from
  exponential electrodynamics}},  {\em Phys. Rev. D} {\bf 113} (2026), no.~8
  084064 [\href{http://arXiv.org/abs/2601.18967}{{\tt 2601.18967}}].

\bibitem{lovelock1970divergence}
D.~Lovelock, {\it Divergence-free tensorial concomitants},  {\em aequationes
  mathematicae} {\bf 4} (1970), no.~1 127--138.

\bibitem{Lovelock:1971yv}
D.~Lovelock, {\it {The Einstein tensor and its generalizations}},  {\em J.
  Math. Phys.} {\bf 12} (1971) 498--501.

\bibitem{Oliva:2010zd}
J.~Oliva and S.~Ray, {\it {Classification of Six Derivative Lagrangians of
  Gravity and Static Spherically Symmetric Solutions}},  {\em Phys. Rev.} {\bf
  D82} (2010) 124030 [\href{http://arXiv.org/abs/1004.0737}{{\tt 1004.0737}}].

\bibitem{Myers:2010ru}
R.~C. Myers and B.~Robinson, {\it {Black Holes in Quasi-topological Gravity}},
  {\em JHEP} {\bf 08} (2010) 067 [\href{http://arXiv.org/abs/1003.5357}{{\tt
  1003.5357}}].

\bibitem{Bueno:2025qjk}
P.~Bueno, R.~A. Hennigar and {\'A}.~J. Murcia, {\it {Birkhoff implies
  quasi-topological}},  {\em Class. Quant. Grav.} {\bf 43} (2026), no.~9 095020
  [\href{http://arXiv.org/abs/2510.25823}{{\tt 2510.25823}}].

\bibitem{Dehghani:2011vu}
M.~H. Dehghani, A.~Bazrafshan, R.~B. Mann, M.~R. Mehdizadeh, M.~Ghanaatian and
  M.~H. Vahidinia, {\it {Black Holes in Quartic Quasitopological Gravity}},
  {\em Phys. Rev.} {\bf D85} (2012) 104009
  [\href{http://arXiv.org/abs/1109.4708}{{\tt 1109.4708}}].

\bibitem{Cisterna:2017umf}
A.~Cisterna, L.~Guajardo, M.~Hassaine and J.~Oliva, {\it {Quintic
  quasi-topological gravity}},  {\em JHEP} {\bf 04} (2017) 066
  [\href{http://arXiv.org/abs/1702.04676}{{\tt 1702.04676}}].

\bibitem{Ahmed:2017jod}
J.~Ahmed, R.~A. Hennigar, R.~B. Mann and M.~Mir, {\it {Quintessential Quartic
  Quasi-topological Quartet}},  {\em JHEP} {\bf 05} (2017) 134
  [\href{http://arXiv.org/abs/1703.11007}{{\tt 1703.11007}}].

\bibitem{Bueno:2019ycr}
P.~Bueno, P.~A. Cano and R.~A. Hennigar, {\it {(Generalized) quasi-topological
  gravities at all orders}},  {\em Class. Quant. Grav.} {\bf 37} (2020), no.~1
  015002 [\href{http://arXiv.org/abs/1909.07983}{{\tt 1909.07983}}].

\bibitem{Bueno:2022res}
P.~Bueno, P.~A. Cano, R.~A. Hennigar, M.~Lu and J.~Moreno, {\it {Generalized
  quasi-topological gravities: the whole shebang}},  {\em Class. Quant. Grav.}
  {\bf 40} (2023), no.~1 015004 [\href{http://arXiv.org/abs/2203.05589}{{\tt
  2203.05589}}].

\bibitem{Moreno:2023rfl}
J.~Moreno and A.~J. Murcia, {\it {Classification of generalized
  quasitopological gravities}},  {\em Phys. Rev. D} {\bf 108} (2023), no.~4
  044016 [\href{http://arXiv.org/abs/2304.08510}{{\tt 2304.08510}}].

\bibitem{Moreno:2023arp}
J.~Moreno and A.~J. Murcia, {\it {Cosmological higher-curvature gravities}},
  {\em Class. Quant. Grav.} {\bf 41} (2024), no.~13 135017
  [\href{http://arXiv.org/abs/2311.12104}{{\tt 2311.12104}}].

\bibitem{Bueno:2024dgm}
P.~Bueno, P.~A. Cano and R.~A. Hennigar, {\it {Regular black holes from pure
  gravity}},  {\em Phys. Lett. B} {\bf 861} (2025) 139260
  [\href{http://arXiv.org/abs/2403.04827}{{\tt 2403.04827}}].

\bibitem{Bueno:2025zaj}
P.~Bueno, P.~A. Cano, R.~A. Hennigar and {\'A}.~J. Murcia, {\it {Regular black
  hole formation in four-dimensional nonpolynomial gravities}},  {\em Phys.
  Rev. D} {\bf 113} (2026), no.~2 024019
  [\href{http://arXiv.org/abs/2509.19016}{{\tt 2509.19016}}].

\bibitem{Aguayo:2025xfi}
M.~Aguayo, L.~Gajardo, N.~Grandi, J.~Moreno, J.~Oliva and M.~Reyes, {\it
  {Holographic explorations of regular black holes in pure gravity}},  {\em
  JHEP} {\bf 09} (2025) 030 [\href{http://arXiv.org/abs/2505.11736}{{\tt
  2505.11736}}].

\bibitem{Hennigar:2025yqm}
R.~A. Hennigar, D.~Kubiz{\v{n}}{\'a}k, S.~Murk and I.~Soranidis, {\it
  {Thermodynamics of regular black holes in anti-de Sitter space}},  {\em JHEP}
  {\bf 11} (2025) 121 [\href{http://arXiv.org/abs/2505.11623}{{\tt
  2505.11623}}].

\bibitem{Borissova:2026wmn}
J.~Borissova and R.~Carballo-Rubio, {\it {Regular black holes from pure gravity
  in four dimensions}},  {\em Phys. Rev. D} {\bf 113} (2026), no.~12 124004
  [\href{http://arXiv.org/abs/2602.16773}{{\tt 2602.16773}}].

\bibitem{Bueno:2026oyg}
P.~Bueno, P.~A. Cano, R.~A. Hennigar and {\'A}.~J. Murcia, {\it {Regular Black
  Holes in Nonlocal Quasitopological Gravity}},
  \href{http://arXiv.org/abs/2607.07790}{{\tt 2607.07790}}.

\bibitem{Cano:2020ezi}
P.~A. Cano and {\'A}.~Murcia, {\it {Resolution of Reissner-Nordstr\"om
  singularities by higher-derivative corrections}},  {\em Class. Quant. Grav.}
  {\bf 38} (2021), no.~7 075014 [\href{http://arXiv.org/abs/2006.15149}{{\tt
  2006.15149}}].

\bibitem{Cano:2020qhy}
P.~A. Cano and {\'A}.~Murcia, {\it {Electromagnetic Quasitopological
  Gravities}},  {\em JHEP} {\bf 10} (2020) 125
  [\href{http://arXiv.org/abs/2007.04331}{{\tt 2007.04331}}].

\bibitem{Bueno:2021krl}
P.~Bueno, P.~A. Cano, J.~Moreno and G.~van~der Velde, {\it {Regular black holes
  in three dimensions}},  {\em Phys. Rev. D} {\bf 104} (2021), no.~2 L021501
  [\href{http://arXiv.org/abs/2104.10172}{{\tt 2104.10172}}].

\bibitem{Bueno:2022ewf}
P.~Bueno, P.~A. Cano, J.~Moreno and G.~van~der Velde, {\it {Electromagnetic
  generalized quasitopological gravities in (2+1) dimensions}},  {\em Phys.
  Rev. D} {\bf 107} (2023), no.~6 064050
  [\href{http://arXiv.org/abs/2212.00637}{{\tt 2212.00637}}].

\bibitem{Bueno:2025dqk}
P.~Bueno, O.~Lasso~Andino, J.~Moreno and G.~van~der Velde, {\it {On regular
  charged black holes in three dimensions}},  {\em JHEP} {\bf 08} (2025) 132
  [\href{http://arXiv.org/abs/2503.02930}{{\tt 2503.02930}}].

\bibitem{Borissova:2026krh}
J.~Borissova, {\it {All 2D generalized dilaton theories from
  d{\ensuremath{\geq}}4 gravities}},  {\em Phys. Rev. D} {\bf 113} (2026),
  no.~12 124088 [\href{http://arXiv.org/abs/2603.06786}{{\tt 2603.06786}}].

\bibitem{Horndeski:1974wa}
G.~W. Horndeski, {\it {Second-order scalar-tensor field equations in a
  four-dimensional space}},  {\em Int. J. Theor. Phys.} {\bf 10} (1974)
  363--384.

\bibitem{Kobayashi:2011nu}
T.~Kobayashi, M.~Yamaguchi and J.~Yokoyama, {\it {Generalized G-inflation:
  Inflation with the most general second-order field equations}},  {\em Prog.
  Theor. Phys.} {\bf 126} (2011) 511--529
  [\href{http://arXiv.org/abs/1105.5723}{{\tt 1105.5723}}].

\bibitem{Kobayashi:2019hrl}
T.~Kobayashi, {\it {Horndeski theory and beyond: a review}},  {\em Rept. Prog.
  Phys.} {\bf 82} (2019), no.~8 086901
  [\href{http://arXiv.org/abs/1901.07183}{{\tt 1901.07183}}].

\bibitem{Carballo-Rubio:2025ntd}
R.~Carballo-Rubio, {\it {Master field equations for spherically symmetric
  gravitational fields beyond general relativity}},  {\em Nature Commun.} {\bf
  17} (2026), no.~1 1399 [\href{http://arXiv.org/abs/2507.15920}{{\tt
  2507.15920}}].

\bibitem{Wald:1993nt}
R.~M. Wald, {\it {Black hole entropy is the Noether charge}},  {\em Phys. Rev.}
  {\bf D48} (1993) 3427--3431 [\href{http://arXiv.org/abs/gr-qc/9307038}{{\tt
  gr-qc/9307038}}].

\bibitem{Iyer:1994ys}
V.~Iyer and R.~M. Wald, {\it {Some properties of Noether charge and a proposal
  for dynamical black hole entropy}},  {\em Phys. Rev.} {\bf D50} (1994)
  846--864 [\href{http://arXiv.org/abs/gr-qc/9403028}{{\tt gr-qc/9403028}}].

\bibitem{Borissova:2026rbi}
J.~Borissova, {\it {$g_{tt}g_{rr} =-1$ black hole thermodynamics in extended
  quasi-topological gravity}},  \href{http://arXiv.org/abs/2604.24101}{{\tt
  2604.24101}}.

\bibitem{Bueno:2024eig}
P.~Bueno, P.~A. Cano, R.~A. Hennigar and A.~J. Murcia, {\it {Dynamical
  Formation of Regular Black Holes}},  {\em Phys. Rev. Lett.} {\bf 134} (2025),
  no.~18 181401 [\href{http://arXiv.org/abs/2412.02742}{{\tt 2412.02742}}].

\bibitem{Bueno:2024zsx}
P.~Bueno, P.~A. Cano, R.~A. Hennigar and A.~J. Murcia, {\it {Regular black
  holes from thin-shell collapse}},  {\em Phys. Rev. D} {\bf 111} (2025),
  no.~10 104009 [\href{http://arXiv.org/abs/2412.02740}{{\tt 2412.02740}}].

\bibitem{Bueno:2025gjg}
P.~Bueno, P.~A. Cano, R.~A. Hennigar, {\'A}.~J. Murcia and A.~Vicente-Cano,
  {\it {Regular black holes from Oppenheimer-Snyder collapse}},  {\em Phys.
  Rev. D} {\bf 112} (2025), no.~6 064039
  [\href{http://arXiv.org/abs/2505.09680}{{\tt 2505.09680}}].

\bibitem{Colleaux:2017ibe}
A.~Coll{\'e}aux, S.~Chinaglia and S.~Zerbini, {\it {Nonpolynomial Lagrangian
  approach to regular black holes}},  {\em Int. J. Mod. Phys. D} {\bf 27}
  (2018), no.~03 1830002 [\href{http://arXiv.org/abs/1712.03730}{{\tt
  1712.03730}}].

\bibitem{Colleaux:2019ckh}
A.~Colleaux, {\em {Regular black hole and cosmological spacetimes in
  Non-Polynomial Gravity theories}}.
\newblock PhD thesis, Trento U., 6, 2019.

\bibitem{Camanho:2010ru}
X.~O. Camanho, J.~D. Edelstein and M.~F. Paulos, {\it {Lovelock theories,
  holography and the fate of the viscosity bound}},  {\em JHEP} {\bf 05} (2011)
  127 [\href{http://arXiv.org/abs/1010.1682}{{\tt 1010.1682}}].

\bibitem{Camanho:2011rj}
X.~O. Camanho and J.~D. Edelstein, {\it {A Lovelock black hole bestiary}},
  {\em Class. Quant. Grav.} {\bf 30} (2013) 035009
  [\href{http://arXiv.org/abs/1103.3669}{{\tt 1103.3669}}].

\bibitem{Camanho:2013pzg}
X.~O. Camanho, {\em {Lovelock gravity, black holes and holography}}.
\newblock PhD thesis, Santiago de Compostela U., 2013.
\newblock \href{http://arXiv.org/abs/1509.08129}{{\tt 1509.08129}}.

\bibitem{Bueno:2019ltp}
P.~Bueno, P.~A. Cano, J.~Moreno and {\'A}.~Murcia, {\it {All higher-curvature
  gravities as Generalized quasi-topological gravities}},  {\em JHEP} {\bf 11}
  (2019) 062 [\href{http://arXiv.org/abs/1906.00987}{{\tt 1906.00987}}].

\bibitem{Padilla:2012ze}
A.~Padilla and V.~Sivanesan, {\it {Boundary Terms and Junction Conditions for
  Generalized Scalar-Tensor Theories}},  {\em JHEP} {\bf 08} (2012) 122
  [\href{http://arXiv.org/abs/1206.1258}{{\tt 1206.1258}}].

\bibitem{Almheiri:2014cka}
A.~Almheiri and J.~Polchinski, {\it {Models of AdS$_{2}$ backreaction and
  holography}},  {\em JHEP} {\bf 11} (2015) 014
  [\href{http://arXiv.org/abs/1402.6334}{{\tt 1402.6334}}].

\bibitem{Maldacena:2016hyu}
J.~Maldacena and D.~Stanford, {\it {Remarks on the Sachdev-Ye-Kitaev model}},
  {\em Phys. Rev. D} {\bf 94} (2016), no.~10 106002
  [\href{http://arXiv.org/abs/1604.07818}{{\tt 1604.07818}}].

\bibitem{Jensen:2016pah}
K.~Jensen, {\it {Chaos in AdS$_2$ Holography}},  {\em Phys. Rev. Lett.} {\bf
  117} (2016), no.~11 111601 [\href{http://arXiv.org/abs/1605.06098}{{\tt
  1605.06098}}].

\bibitem{Engelsoy:2016xyb}
J.~Engels{\"o}y, T.~G. Mertens and H.~Verlinde, {\it {An investigation of
  AdS$_{2}$ backreaction and holography}},  {\em JHEP} {\bf 07} (2016) 139
  [\href{http://arXiv.org/abs/1606.03438}{{\tt 1606.03438}}].

\bibitem{Sarosi:2017ykf}
G.~S{\'a}rosi, {\it {AdS$_{2}$ holography and the SYK model}},  {\em PoS} {\bf
  Modave2017} (2018) 001 [\href{http://arXiv.org/abs/1711.08482}{{\tt
  1711.08482}}].

\bibitem{Alekseev:1988ce}
A.~Alekseev and S.~L. Shatashvili, {\it {Path Integral Quantization of the
  Coadjoint Orbits of the Virasoro Group and 2D Gravity}},  {\em Nucl. Phys. B}
  {\bf 323} (1989) 719--733.

\bibitem{Witten:1987ty}
E.~Witten, {\it {Coadjoint Orbits of the Virasoro Group}},  {\em Commun. Math.
  Phys.} {\bf 114} (1988) 1.

\bibitem{Duistermaat:1982vw}
J.~J. Duistermaat and G.~J. Heckman, {\it {On the Variation in the cohomology
  of the symplectic form of the reduced phase space}},  {\em Invent. Math.}
  {\bf 69} (1982) 259--268.

\bibitem{Moitra:2021uiv}
U.~Moitra, S.~K. Sake and S.~P. Trivedi, {\it {Jackiw-Teitelboim gravity in the
  second order formalism}},  {\em JHEP} {\bf 10} (2021) 204
  [\href{http://arXiv.org/abs/2101.00596}{{\tt 2101.00596}}].

\bibitem{Quine1993}
J.~R. Quine, S.~H. Heydari and R.~Y. Song, {\it Zeta regularized products},
  {\em Transactions of the American Mathematical Society} {\bf 338} (1993),
  no.~1 213--231.

\bibitem{Colleaux:2026hat}
A.~Coll{\'e}aux, {\it {Rational regular black holes in non-polynomial
  gravity}},  \href{http://arXiv.org/abs/2608.17158}{{\tt 2608.17158}}.

\bibitem{Hayward:2005gi}
S.~A. Hayward, {\it {Formation and evaporation of regular black holes}},  {\em
  Phys. Rev. Lett.} {\bf 96} (2006) 031103
  [\href{http://arXiv.org/abs/gr-qc/0506126}{{\tt gr-qc/0506126}}].

\bibitem{Hossenfelder:2009fc}
S.~Hossenfelder, L.~Modesto and I.~Premont-Schwarz, {\it {A Model for
  non-singular black hole collapse and evaporation}},  {\em Phys. Rev. D} {\bf
  81} (2010) 044036 [\href{http://arXiv.org/abs/0912.1823}{{\tt 0912.1823}}].

\bibitem{Frolov:2014jva}
V.~P. Frolov, {\it {Information loss problem and a 'black hole` model with a
  closed apparent horizon}},  {\em JHEP} {\bf 05} (2014) 049
  [\href{http://arXiv.org/abs/1402.5446}{{\tt 1402.5446}}].

\bibitem{Carballo-Rubio:2018pmi}
R.~Carballo-Rubio, F.~Di~Filippo, S.~Liberati, C.~Pacilio and M.~Visser, {\it
  {On the viability of regular black holes}},  {\em JHEP} {\bf 07} (2018) 023
  [\href{http://arXiv.org/abs/1805.02675}{{\tt 1805.02675}}].

\bibitem{Barcelo:2020mjw}
C.~Barcel{\'o}, V.~Boyanov, R.~Carballo-Rubio and L.~J. Garay, {\it {Black hole
  inner horizon evaporation in semiclassical gravity}},  {\em Class. Quant.
  Grav.} {\bf 38} (2021), no.~12 125003
  [\href{http://arXiv.org/abs/2011.07331}{{\tt 2011.07331}}].

\bibitem{Cadoni:2023tse}
M.~Cadoni, M.~Oi and A.~P. Sanna, {\it {Evaporation and information puzzle for
  2D nonsingular asymptotically flat black holes}},  {\em JHEP} {\bf 06} (2023)
  211 [\href{http://arXiv.org/abs/2303.05557}{{\tt 2303.05557}}].

\bibitem{Barenboim:2024dko}
J.~Barenboim, A.~V. Frolov and G.~Kunstatter, {\it {No drama in two-dimensional
  black hole evaporation}},  {\em Phys. Rev. Res.} {\bf 6} (2024), no.~3
  L032055 [\href{http://arXiv.org/abs/2405.13373}{{\tt 2405.13373}}].

\bibitem{Barenboim:2025ckx}
J.~Barenboim, A.~V. Frolov and G.~Kunstatter, {\it {Evaporation of regular
  black holes in 2D dilaton gravity}},  {\em Phys. Rev. D} {\bf 111} (2025),
  no.~10 104068 [\href{http://arXiv.org/abs/2503.03191}{{\tt 2503.03191}}].

\bibitem{Carballo-Rubio:2026gwg}
R.~Carballo-Rubio, F.~Di~Filippo, S.~Liberati and M.~Visser, {\it
  {Semiclassical regularity of compact trapped regions: From dynamical horizons
  to inner extremality}},  \href{http://arXiv.org/abs/2607.03916}{{\tt
  2607.03916}}.

\bibitem{Arrechea:2026rua}
J.~Arrechea, S.~Liberati and M.~Spadafora, {\it {Semiclassical Black Hole-White
  Hole transitions: an analytical treatment}},
  \href{http://arXiv.org/abs/2608.03538}{{\tt 2608.03538}}.

\bibitem{Easson:2026vte}
D.~A. Easson, {\it {Anomaly-driven evaporation endpoints of a two-dimensional
  regular black hole}},  \href{http://arXiv.org/abs/2606.09983}{{\tt
  2606.09983}}.

\bibitem{Brown:2024ajk}
A.~R. Brown, L.~V. Iliesiu, G.~Penington and M.~Usatyuk, {\it {The evaporation
  of charged black holes}},  \href{http://arXiv.org/abs/2411.03447}{{\tt
  2411.03447}}.

\end{thebibliography}\endgroup

\end{document}